\documentclass[12pt]{article}
\usepackage{amsmath,amssymb,amsthm,amsxtra,overpic,bbm,bm,epsfig,subfigure}
\usepackage{hyperref}
\usepackage{graphicx}
\usepackage{color}
\usepackage{comment}
\usepackage{epstopdf}
\usepackage{cite}
\usepackage{hyperref}
\usepackage{slashed,stmaryrd}

\usepackage{lscape}
\usepackage{array}
\usepackage{booktabs}

\def\thefootnote{\fnsymbol{footnote}}

\begin{document}

\vspace{0.2cm}

\begin{center}
{\Large\bf Analytical solutions to renormalization-group equations of effective neutrino masses and mixing parameters in matter}
\end{center}

\vspace{0.2cm}

\begin{center}
{\bf Xin Wang} \footnote{E-mail: wangx@ihep.ac.cn},
\quad
{\bf Shun Zhou} \footnote{E-mail: zhoush@ihep.ac.cn}
\\
{\small Institute of High Energy Physics, Chinese Academy of
Sciences, Beijing 100049, China \\
School of Physical Sciences, University of Chinese Academy of Sciences, Beijing 100049, China}
\end{center}

\vspace{1.5cm}

\begin{abstract}
Neutrino oscillations in matter can be fully described by six effective parameters, namely, three neutrino mixing angles $\{\widetilde{\theta}^{}_{12}, \widetilde{\theta}^{}_{13}, \widetilde{\theta}^{}_{23}\}$, one Dirac-type CP-violating phase $\widetilde{\delta}$, and two neutrino mass-squared differences $\widetilde{\Delta}^{}_{21} \equiv \widetilde{m}^2_2 - \widetilde{m}^2_1$ and $\widetilde{\Delta}^{}_{31} \equiv \widetilde{m}^2_3 - \widetilde{m}^2_1$. Recently, a complete set of differential equations for these effective parameters have been derived to characterize their evolution with respect to the ordinary matter term $a \equiv 2\sqrt{2}G^{}_{\rm F} N^{}_e E$, in analogy with the renormalization-group equations (RGEs) for running parameters. Via series expansion in terms of the small ratio $\alpha^{}_{\rm c} \equiv \Delta^{}_{21}/\Delta^{}_{\rm c}$ with $\Delta^{}_{\rm c} \equiv \Delta^{}_{31} \cos^2 \theta^{}_{12} + \Delta^{}_{32} \sin^2 \theta^{}_{12}$, we obtain approximate analytical solutions to the RGEs of the effective neutrino parameters and make several interesting observations. First, at the leading order, $\widetilde{\theta}^{}_{12}$ and $\widetilde{\theta}^{}_{13}$ are given by the simple formulas in the two-flavor mixing limit, while $\widetilde{\theta}^{}_{23} \approx \theta^{}_{23}$ and $\widetilde{\delta} \approx \delta$ are not changed by matter effects. Second, the ratio of the matter-corrected Jarlskog invariant $\widetilde{\cal J}$ to its counterpart in vacuum ${\cal J}$ approximates to $\widetilde{\cal J}/{\cal J} \approx 1/(\widehat{C}^{}_{12} \widehat{C}^{}_{13})$, where $\widehat{C}^{}_{12} \equiv \sqrt{1 - 2 A^{}_* \cos 2\theta^{}_{12} + A^2_*}$ with $A^{}_* \equiv a/\Delta^{}_{21}$ and $\widehat{C}^{}_{13} \equiv \sqrt{1 - 2 A^{}_{\rm c} \cos 2\theta^{}_{13} + A^2_{\rm c}}$ with $A^{}_{\rm c} \equiv a/\Delta^{}_{\rm c}$ have been defined. Finally, after taking higher-order corrections into account, we find compact and simple expressions of all the effective parameters, which turn out to be in perfect agreement with the exact numerical results.
\end{abstract}


\def\thefootnote{\arabic{footnote}}
\setcounter{footnote}{0}

\newpage
\section{Introduction}
The matter effects on neutrino oscillations have been playing an important role in understanding various neutrino oscillation data, in particular those from the solar, accelerator and atmospheric neutrino experiments~\cite{Wolfenstein:1977ue, Mikheev:1986gs, Mikheyev:1989dy, Kuo:1989qe}. In the framework of three-flavor neutrino mixing, neutrino oscillations in ordinary matter are governed by the effective Hamiltonian
\begin{eqnarray}
H^{}_{\rm eff} = \frac{1}{2E}\left[ U \left(\begin{matrix} m^2_1 & 0 & 0 \cr 0 & m^2_2 & 0 \cr 0 & 0 & m^2_3 \end{matrix}\right) U^\dagger + \left(\begin{matrix} a & 0 & 0 \cr 0 & 0 & 0 \cr 0 & 0 & 0 \end{matrix}\right)\right] \; ,
\label{eq:Heff1}
\end{eqnarray}
where $E$ is the neutrino beam energy, and $a \equiv 2\sqrt{2} G^{}_{\rm F} N^{}_e E$ is the matter parameter with $G^{}_{\rm F}$ and $N^{}_e$ being the Fermi constant and the net electron number density, respectively. In Eq.~(\ref{eq:Heff1}), $m^{}_i$ (for $i = 1, 2, 3$) stand for neutrino masses in vacuum, and $U$ is the unitary lepton flavor mixing matrix in vacuum. After diagonalizing the effective Hamiltonian via the unitary transformation
\begin{eqnarray}
H^{}_{\rm eff} = \frac{1}{2E} \left[ V \left(\begin{matrix} \widetilde{m}^2_1 & 0 & 0 \cr 0 & \widetilde{m}^2_2 & 0 \cr 0 & 0 & \widetilde{m}^2_3 \end{matrix}\right) V^\dagger \right] \; ,
\label{eq:Heff2}
\end{eqnarray}
where $\widetilde{m}^{}_i$ (for $i = 1, 2, 3$) denote the effective neutrino masses and $V$ is the effective neutrino mixing matrix in matter, one can easily calculate the oscillation probabilities by using the effective mixing parameters in the same way as in the case of neutrino oscillations in vacuum.

Recently, it has been shown in Refs.~\cite{Chiu:2010da, Chiu:2017ckv, Xing:2018} that the elements of the effective neutrino mixing matrix $V^{}_{\alpha i}$ (for $\alpha = e, \mu, \tau$ and $i = 1, 2, 3$) and the effective neutrino mass-squared differences $\widetilde{\Delta}^{}_{ij} \equiv \widetilde{m}^2_i - \widetilde{m}^2_j$ (for $ij = 21, 31, 32$) satisfy a complete set of differential equations with respect to the matter parameter $a$. Making an analogy with the renormalization-group equations (RGEs) and adopting the standard parametrization for the effective mixing matrix $V$ in matter, the authors of Ref.~\cite{Xing:2018} have derived the RGEs for the effective mixing angles $\{\widetilde{\theta}^{}_{12}, \widetilde{\theta}^{}_{13}, \widetilde{\theta}^{}_{23}\}$ and the effective CP-violating phase $\widetilde{\delta}$, namely,
\begin{eqnarray}
\frac{{\rm d}\widetilde{\theta}^{}_{12}}{{\rm d}a}  &=& \dfrac{1}{2} \sin 2\widetilde{\theta}^{}_{12} \left( \cos^2 \widetilde{\theta}^{}_{13} \widetilde{\Delta}_{21}^{-1} - \sin^2 \widetilde{\theta}^{}_{13} \widetilde{\Delta}_{21}^{} \widetilde{\Delta}_{31}^{-1} \widetilde{\Delta}_{32}^{-1} \right) \; , \label{eq:rgetheta12}\\
\frac{{\rm d}\widetilde{\theta}^{}_{13}}{{\rm d}a}  &=& \frac{1}{2} \sin 2\widetilde{\theta}^{}_{13} \left( \cos^2 \widetilde{\theta}^{}_{12} \widetilde{\Delta}_{31}^{-1} + \sin^2 \widetilde{\theta}_{12}^{} \widetilde{\Delta}_{32}^{-1} \right) \; , \label{eq:rgetheta13}\\
\frac{{\rm d}\widetilde{\theta}^{}_{23}}{{\rm d}a}  &=& \frac{1}{2} \sin 2\widetilde{\theta}^{}_{12} \sin \widetilde{\theta}^{}_{13} \cos \widetilde{\delta} \widetilde{\Delta}^{}_{21} \widetilde{\Delta}_{31}^{-1} \widetilde{\Delta}_{32}^{-1} \; , \label{eq:rgetheta23} \\
\frac{{\rm d}\widetilde{\delta}}{{\rm d}a}  &=& -\sin 2\widetilde{\theta}^{}_{12} \sin \widetilde{\theta}^{}_{13} \sin \widetilde{\delta} \cot 2\widetilde{\theta}^{}_{23} \widetilde{\Delta}^{}_{21} \widetilde{\Delta}_{31}^{-1} \widetilde{\Delta}_{32}^{-1} \; ; \label{eq:rgedelta}
\end{eqnarray}
as well as the RGEs for the effective neutrino mass-squared differences $\{\widetilde{\Delta}^{}_{21}, \widetilde{\Delta}^{}_{31}, \widetilde{\Delta}^{}_{32}\}$, i.e.,
\begin{eqnarray}
\frac{\mathrm{d} \widetilde{\Delta}_{21}}{\mathrm{d}a} &=& -\cos^2 \widetilde{\theta}^{}_{13} \cos 2\widetilde{\theta}^{}_{12} \; , \label{eq:rgeDel21} \\
\frac{\mathrm{d} \widetilde{\Delta}_{31}}{\mathrm{d}a} &=& \sin^2 \widetilde{\theta}^{}_{13} - \cos^2 \widetilde{\theta}^{}_{13} \cos^2 \widetilde{\theta}^{}_{12} \; , \label{eq:rgeDel31} \\
\frac{\mathrm{d}\widetilde{\Delta}_{32}}{\mathrm{d}a} &=& \sin^2 \widetilde{\theta}^{}_{13} - \cos^2 \widetilde{\theta}^{}_{13} \sin^2 \widetilde{\theta}^{}_{12} \; , \label{eq:rgeDel32}
\end{eqnarray}
where it is evident that only two of the above three equations are independent. These RGEs have been exactly solved in Ref.~\cite{Xing:2018} in a numerical way, which however obscures how exactly the matter effects modify the effective neutrino masses and mixing parameters.

In this paper, we present the first analytical solutions to those RGEs with some reasonable approximations and compare them with the exact numerical results. Such a comparison is very helpful for us to understand how the matter effects change the effective parameters and thus the oscillation probabilities. Very interestingly, it has been observed in Refs.~\cite{Denton:2016wmg, Denton:2018hal, Ioannisian:2018qwl, Denton:2018fex} that the effective mixing angles $\widetilde{\theta}^{}_{12}$ and $\widetilde{\theta}^{}_{13}$ are approximately given by the simple formulas in the two-flavor mixing limit. As we shall see shortly, this observation follows naturally as the leading-order analytical solutions to the RGEs of the effective mixing angles. Moreover, we derive the approximate analytical expressions for all the effective neutrino mass-squared differences and mixing angles.

The remaining part of this paper is structured as follows. In Sec. 2, we briefly summarize the series expansion of effective neutrino mass-squared differences in terms of the perturbation parameter $\alpha \equiv \Delta^{}_{21}/\Delta^{}_{31}$, where $\Delta^{}_{21} \equiv m^2_2 - m^2_1 \approx 7.5\times 10^{-5}~{\rm eV}^2$ and $|\Delta^{}_{31}| \equiv |m^2_3 - m^2_1| \approx 2.5\times 10^{-3}~{\rm eV}^2$ are the neutrino mass-squared differences in vacuum. The relevant results serve as the starting point for the analytical solutions to the RGEs. Then, the analytical results are derived in Sec. 3, and compared with the exact numerical solutions. Finally, we make some concluding remarks in Sec. 4.

\section{Series Expansion}

In fact, the eigenvalues and eigenvectors of the effective Hamiltonian in Eq.~(\ref{eq:Heff1}) can be exactly calculated and expressed in terms of neutrino masses $\{m^{}_1, m^{}_2, m^{}_3\}$ and the mixing parameters $\{\theta^{}_{12}, \theta^{}_{13}, \theta^{}_{23}, \delta\}$ in vacuum, if the standard parametrization of the mixing matrix is adopted, and the matter parameter $a$~\cite{Zaglauer:1988, Xing:2001, Xing:2004}. Therefore, the oscillation probabilities can be computed by using the effective neutrino mass eigenvalues and mixing matrix. Nevertheless, it is difficult to directly confront the exact oscillation probabilities with neutrino oscillation data in order to figure out how the matter effects change the oscillation behavior. In Refs.~\cite{Cervera:2000kp, Freund:2001, Akhmedov2004}, the analytical formulas of effective neutrino masses and mixing parameters have been derived via series expansion with respect to $\alpha = \Delta^{}_{21}/\Delta^{}_{31} \approx 0.03$ or $\sin^2 \theta^{}_{13} \approx 0.02$ or both. Then neutrino oscillation probabilities can be computed and implemented to understand those neutrino oscillation experiments where matter effects play a significant role.

For our purpose, we follow the same formalism in the seminal paper by Freund~\cite{Freund:2001} and quote the results of the effective neutrino mass-squared differences to the first order of $\alpha$ as below
\begin{eqnarray}
\widetilde{\Delta}^{}_{21} &\approx& \Delta^{}_{31} \left[\frac{1}{2} \left(1 + A - C^{}_{13}\right) + \alpha \left(\frac{C^{}_{13} + 1 - A \cos 2\theta^{}_{13}}{2C^{}_{13}} \sin^2 \theta^{}_{12} - \cos^2 \theta^{}_{12} \right)\right] \; , \label{eq:Del21} \\
\widetilde{\Delta}_{31} &\approx& \Delta^{}_{31} \left[\frac{1}{2} \left(1 + A + C^{}_{13}\right) + \alpha \left(\frac{C^{}_{13} - 1 + A \cos 2\theta^{}_{13}}{2C^{}_{13}} \sin^2 \theta^{}_{12} - \cos^2 \theta^{}_{12}\right)\right] \; , \label{eq:Del31} \\
\widetilde{\Delta}_{32} &\approx& \Delta^{}_{31} \left[ C^{}_{13} + \alpha \sin^2 \theta^{}_{12} \left(\frac{A \cos 2\theta^{}_{13} - 1}{C^{}_{13}} \right)\right] \; , \label{eq:Del32}
\end{eqnarray}
where $A \equiv a/\Delta^{}_{31}$ and $C^{}_{13} \equiv \sqrt{1 - 2A \cos 2\theta^{}_{13}+ A^2}$ have been defined, and the higher-order terms ${\cal O}(\alpha^2)$ have been safely omitted for $A > \alpha$. It has been pointed out in Ref.~\cite{Zhou2017} that the effective Hamiltonian for neutrino oscillations in matter is intrinsically invariant under the transformations $\theta^{}_{12} \rightarrow \theta^{}_{12} - \pi/2$ for the mixing angle and $m^{}_1 \leftrightarrow m^{}_2$ for neutrino masses when one takes the standard parametrization of the mixing matrix in vacuum. To preserve such a symmetry of the effective Hamiltonian in the series expansion at each order, we can introduce a special neutrino mass-squared difference $\Delta^{}_{\rm c} \equiv \Delta^{}_{31} \cos^2 \theta^{}_{12} + \Delta^{}_{32} \sin^2 \theta^{}_{12}$, which has been demonstrated to be the most favorable choice to achieve compact expressions for the effective mixing parameters as well as neutrino oscillation probabilities in matter~\cite{Minakata:2015gra, Parke:2016joa, Li:2016pzm}. After converting into such a symmetric formalism and carrying out the series expansion in terms of $\alpha^{}_{\rm c} \equiv \Delta^{}_{21}/\Delta^{}_{\rm c}$, one can find that the effective neutrino mass-squared differences in matter in Eqs.~(\ref{eq:Del21})-(\ref{eq:Del32}) can be greatly simplified
\begin{eqnarray}
\widetilde{\Delta}^{}_{21} &\approx& \Delta^{}_{\rm c} \left[\frac{1}{2}\left(1 + A^{}_{\rm c} - \widehat{C}^{}_{13}\right) - \alpha^{}_{\rm c} \cos 2\theta^{}_{12} \right] \; , \label{eq:Del21c} \\
\widetilde{\Delta}^{}_{31} &\approx& \Delta^{}_{\rm c} \left[\frac{1}{2}\left(1 + A^{}_{\rm c} + \widehat{C}^{}_{13}\right) - \alpha^{}_{\rm c} \cos 2\theta^{}_{12} \right] \; , \label{eq:Del31c} \\
\widetilde{\Delta}^{}_{32} &\approx& \Delta^{}_{\rm c} \widehat{C}^{}_{13} \; , \label{eq:Del32c}
\end{eqnarray}
where $A^{}_{\rm c} \equiv a/\Delta^{}_{\rm c}$ and $\widehat{C}^{}_{13} \equiv \sqrt{1 - 2 A^{}_{\rm c} \cos 2\theta^{}_{13}+ A^2_{\rm c}} = \sqrt{(A^{}_{\rm c} - \cos 2\theta^{}_{13})^2 + \sin^2 2\theta^{}_{13}}$ are defined. It is straightforward to verify that the expressions on the right-hand sides of Eqs.~(\ref{eq:Del21c})-(\ref{eq:Del32c}) are invariant under the replacements $\cos 2\theta^{}_{12} \to -\cos 2\theta^{}_{12}$ and $\alpha^{}_{\rm c} \to -\alpha^{}_{\rm c}$, which are implied by the transformations $\theta^{}_{12} \to \theta^{}_{12} - \pi/2$ and $m^{}_1 \leftrightarrow m^{}_2$. Moreover, the first-order terms in Eqs.~(\ref{eq:Del21c})-(\ref{eq:Del32c}) become much simper than those in Eqs.~(\ref{eq:Del21})-(\ref{eq:Del32}). In particular, only the zeroth-order term in Eq.~(\ref{eq:Del32c}) survives. Such a considerable simplification will help us a lot to find the analytical solutions to the RGEs. For this reason, we shall use the notations and results in Eqs.~(\ref{eq:Del21c})-(\ref{eq:Del32c}) in the following discussions. As the expressions of effective neutrino mass-squared differences in Eqs.~(\ref{eq:Del21c})-(\ref{eq:Del32c}) are much simpler than those in Eqs.~(\ref{eq:Del21})-(\ref{eq:Del32}), it is also natural to expect that more compact formulas of neutrino oscillation probabilities in the former case can be derived. For instance, in the series expansion of neutrino oscillation probabilities in terms of $\alpha^{}_{\rm c}$, all the terms proportional to $\sin^2_{}\theta^{}_{12}$ or $\cos^2 \theta^{}_{12}$ will disappear. This observation has been demonstrated by explicit calculations of neutrino oscillation probabilities~\cite{WZ}.

Some comments on the series expansion of $\widetilde{\Delta}^{}_{ij}$ (for $ij = 21, 31, 32$) and the mixing parameters should be useful. As has been already stated in Ref.~\cite{Freund:2001}, the series expansion with respect to $\alpha^{}_{\rm c} = \Delta^{}_{21}/\Delta^{}_{\rm c}$ cannot be valid for an arbitrary value of the matter parameter $a$. First of all, the results are no longer correct in the vacuum limit with $a = 0$, as $A^{}_{\rm c} = a/\Delta^{}_{\rm c}$ is always assumed to be nonzero in Ref.~\cite{Freund:2001}~\footnote{Although the notations in the present work are slightly different from those in Ref.~\cite{Freund:2001}, one can easily identify that the relevant quantities $\{a, \Delta^{}_{\rm c}, \alpha^{}_{\rm c}, A^{}_{\rm c}\}$ correspond to $\{A, \Delta, \alpha, \widehat{A}\}$ therein, respectively. Here we just quote the statements from Ref.~\cite{Freund:2001} and translate them into our own notations.}. Second, the series expansion works well only for a relatively large $A^{}_{\rm c}$, e.g., $A^{}_{\rm c} > \alpha^{}_{\rm c}$, corresponding to $E \gtrsim 0.4~{\rm GeV}$ in the case of $\Delta^{}_{21} = 7.5\times 10^{-5}~{\rm eV}^2$ and the matter density $\rho = 2.8~{\rm g}~{\rm cm}^{-3}$. This is why one gets the divergence $\widetilde{\theta}^{}_{12} \to \infty$ in the limit of $A^{}_{\rm c} \to 0$. For a small value $A^{}_{\rm c} \lesssim \alpha^{}_{\rm c}$, one can certainly find another suitable form of series expansion~\cite{Xing:2016ymg, Xu:2015kma}, which is however invalid for a larger value of $A^{}_{\rm c}$. The main reason is that there may exist two resonances at $A^{}_{\rm c} = \alpha^{}_{\rm c} \cos 2\theta^{}_{12}$ and $A^{}_{\rm c} = \cos 2\theta^{}_{13}$, and the level crossing of two relevant eigenvalues occurs at each resonance. The series expansion has to focus on one resonance, and thus cannot be utilized to fully and correctly describe the effective neutrino masses and mixing parameters the whole range of $a$ or equivalently $A^{}_{\rm c}$.

In the next section, starting with the series expansion of $\widetilde{\Delta}^{}_{ij}$ (for $ij = 21, 31, 32$) in Eqs.~(\ref{eq:Del21c})-(\ref{eq:Del32c}), we try to solve the RGEs for effective neutrino mixing angles $\{\widetilde{\theta}^{}_{12}, \widetilde{\theta}^{}_{13}, \widetilde{\theta}^{}_{23}\}$ and the CP violating phase $\widetilde{\delta}$. In order to find out a meaningful solution to $\widetilde{\theta}^{}_{12}$, we are forced to generalize the expansion in Eqs.~(\ref{eq:Del21c})-(\ref{eq:Del32c}) by modifying the first-order terms. As will be shown later, such a modification is not arbitrary but will be regularized by the RGEs, which should be fulfilled strictly no matter whether $A^{}_{\rm c}$ is large or small.

\section{Analytical Solutions}

Now we are ready to analytically solve the RGEs. It is worthwhile to stress that the RGEs for $\{\widetilde{\theta}^{}_{12}, \widetilde{\theta}^{}_{13}\}$ and $\{\widetilde{\Delta}^{}_{21}, \widetilde{\Delta}^{}_{31}, \widetilde{\Delta}^{}_{32}\}$ form a closed set of differential equations~\cite{Xing:2018}, so one can first look for the solutions to those parameters and then go further to solve $\widetilde{\theta}^{}_{23}$ and $\widetilde{\delta}$. For clarity, let us first concentrate on neutrino oscillations in the case of normal neutrino mass ordering (NO), namely, $m^2_3 > m^2_2 > m^2_1$ or $\Delta^{}_{\rm c} > 0$, and then turn to the case of inverted neutrino mass ordering (IO), namely, $m^2_3 < m^2_1 < m^2_2$ or $\Delta^{}_{\rm c} < 0$ later.

\subsection{$\widetilde{\theta}^{}_{13}$ and $\widetilde{\theta}^{}_{12}$}
The first step is to take the series expansion in Eqs.~(\ref{eq:Del21c})-(\ref{eq:Del32c}) as the approximate solutions to $\widetilde{\Delta}^{}_{ij}$ (for $ij = 21, 31, 32$) and insert these solutions into their RGEs in Eqs.~(\ref{eq:rgeDel21})-(\ref{eq:rgeDel32}). After doing so, one can observe from Eqs.~(\ref{eq:rgeDel31}) and (\ref{eq:rgeDel32}) that
\begin{eqnarray}
\frac{1}{2} \left( 1 + \frac{A^{}_{\rm c} - \cos 2\theta^{}_{13}}{\widehat{C}^{}_{13}} \right) &=& \sin^2 \widetilde{\theta}^{}_{13} - \cos^2 \widetilde{\theta}^{}_{13} \cos^2 \widetilde{\theta}^{}_{12} \; , \label{eq:solth13a}\\
\frac{A^{}_{\rm c} - \cos 2\theta^{}_{13}}{\widehat{C}^{}_{13}} &=& \sin^2 \widetilde{\theta}^{}_{13} - \cos^2 \widetilde{\theta}^{}_{13} \sin^2 \widetilde{\theta}^{}_{12} \; , \label{eq:solth13b}
\end{eqnarray}
where the derivative ${\rm d}\widehat{C}^{}_{13}/{\rm d}A^{}_{\rm c} = (A^{}_{\rm c} - \cos 2\theta^{}_{13})/\widehat{C}^{}_{13}$ has been used. It is straightforward to solve $\widetilde{\theta}^{}_{13}$ by adding Eq.~(\ref{eq:solth13a}) to Eq.~(\ref{eq:solth13b}) on both left-hand and right-hand sides, i.e.,
\begin{eqnarray}
\cos^2 \widetilde{\theta}^{}_{13} = \frac{1}{2} \left( 1 - \frac{A^{}_{\rm c} - \cos 2\theta^{}_{13}}{\widehat{C}^{}_{13}} \right) \; ,
\label{eq:solth13no}
\end{eqnarray}
which is the well-known effective mixing angle in the limit of two-flavor neutrino mixing with $\theta^{}_{13}$ being the mixing angle in vacuum and $\Delta^{}_{\rm c}$ being the neutrino mass-squared difference in vacuum. To see this point clearly, we can recast Eq.~(\ref{eq:solth13no}) into a more familiar form
\begin{eqnarray}
\sin^2 2\widetilde{\theta}^{}_{13} = 1 - \frac{(A^{}_{\rm c} - \cos 2\theta^{}_{13})^2}{\widehat{C}^2_{13}} = \frac{\sin^2 2\theta^{}_{13}}{(A^{}_{\rm c} - \cos 2\theta^{}_{13})^2 + \sin^2 2\theta^{}_{13}} \; ,
\label{eq:solth13new}
\end{eqnarray}
implying the maximal effective mixing angle $\widetilde{\theta}^{}_{13} = 45^\circ$ at the resonance $A^{}_{\rm c} = \cos 2\theta^{}_{13}$. For antineutrino oscillations, the solution to $\widetilde{\theta}^{}_{13}$ can be obtained by setting $A^{}_{\rm c} \to - A^{}_{\rm c}$ in Eq.~(\ref{eq:solth13no}). In addition, one can immediately verify that Eq.~(\ref{eq:solth13no}) or Eq.~(\ref{eq:solth13new}) leads to $\widetilde{\theta}^{}_{13} \to \theta^{}_{13}$ in the vacuum limit $A^{}_{\rm c} \to 0$, indicating the correct asymptotic behavior even though the series expansion is in principle valid only for $A^{}_{\rm c} > \alpha^{}_{\rm c}$. This observation can be understood as follows. As was previously noticed in Refs.~\cite{Xing:2016ymg, Xu:2015kma, Li:2016pzm}, it is inappropriate to expand the functions like $\sqrt{A^2_{\rm c} + \alpha^2_{\rm c}}$ with respect to $\alpha^{}_{\rm c}$ in the presence of a comparable or even smaller $A^{}_{\rm c}$. It has also been demonstrated in Ref.~\cite{Xu:2015kma} that such a problem can be avoided when one makes use of the series expansion of $\widetilde{m}^2_2 + \widetilde{m}^2_1$ but not that of $\widetilde{m}^2_2 - \widetilde{m}^2_1$. This is obviously the case for the solution to $\widetilde{\theta}^{}_{13}$ in Eq.~(\ref{eq:solth13no}), which has been derived from Eqs.~(\ref{eq:solth13a}) and (\ref{eq:solth13b}).
\begin{figure}[t!]
\centering
\includegraphics[width=0.49\textwidth]{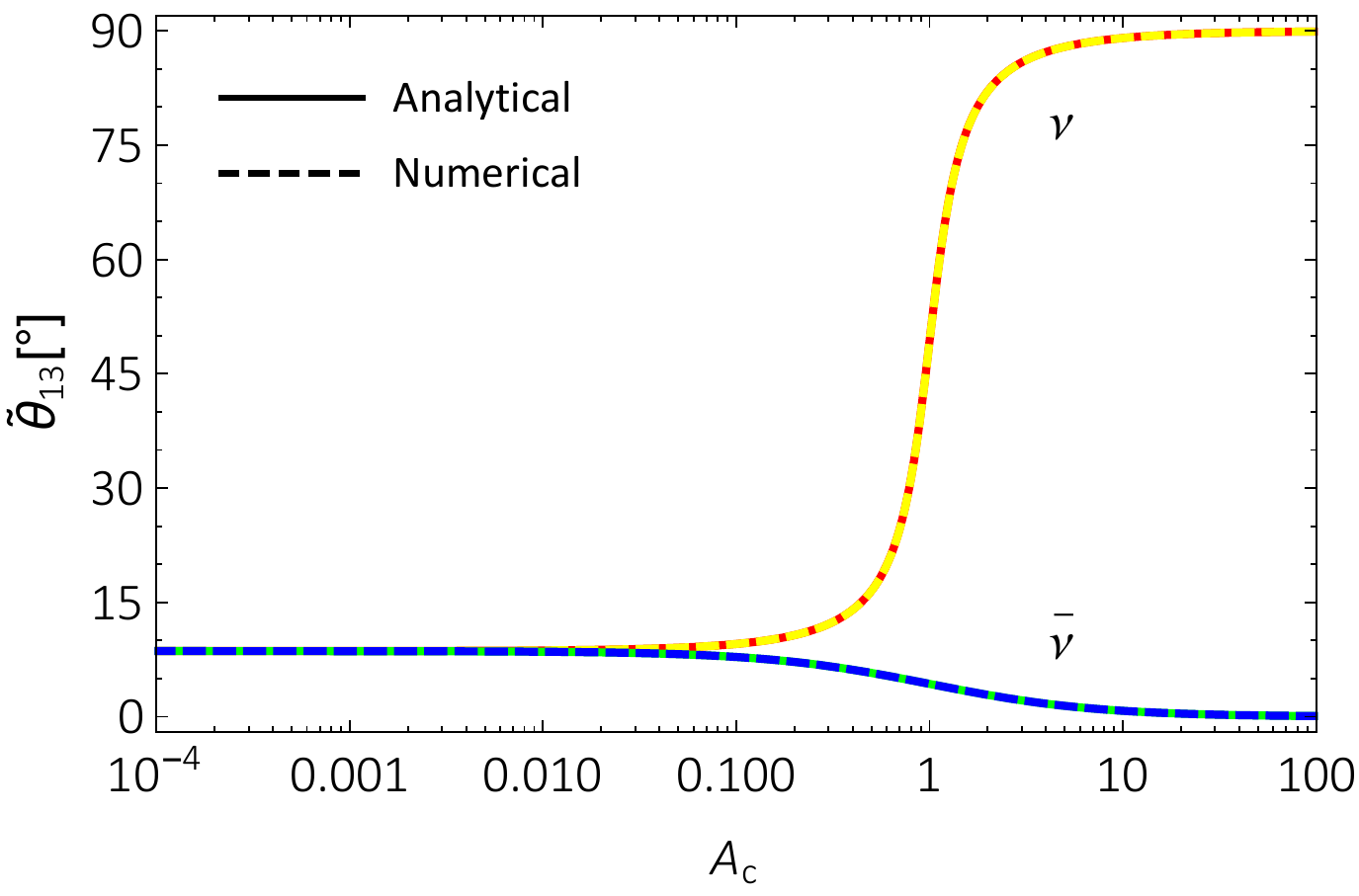}
\includegraphics[width=0.498\textwidth]{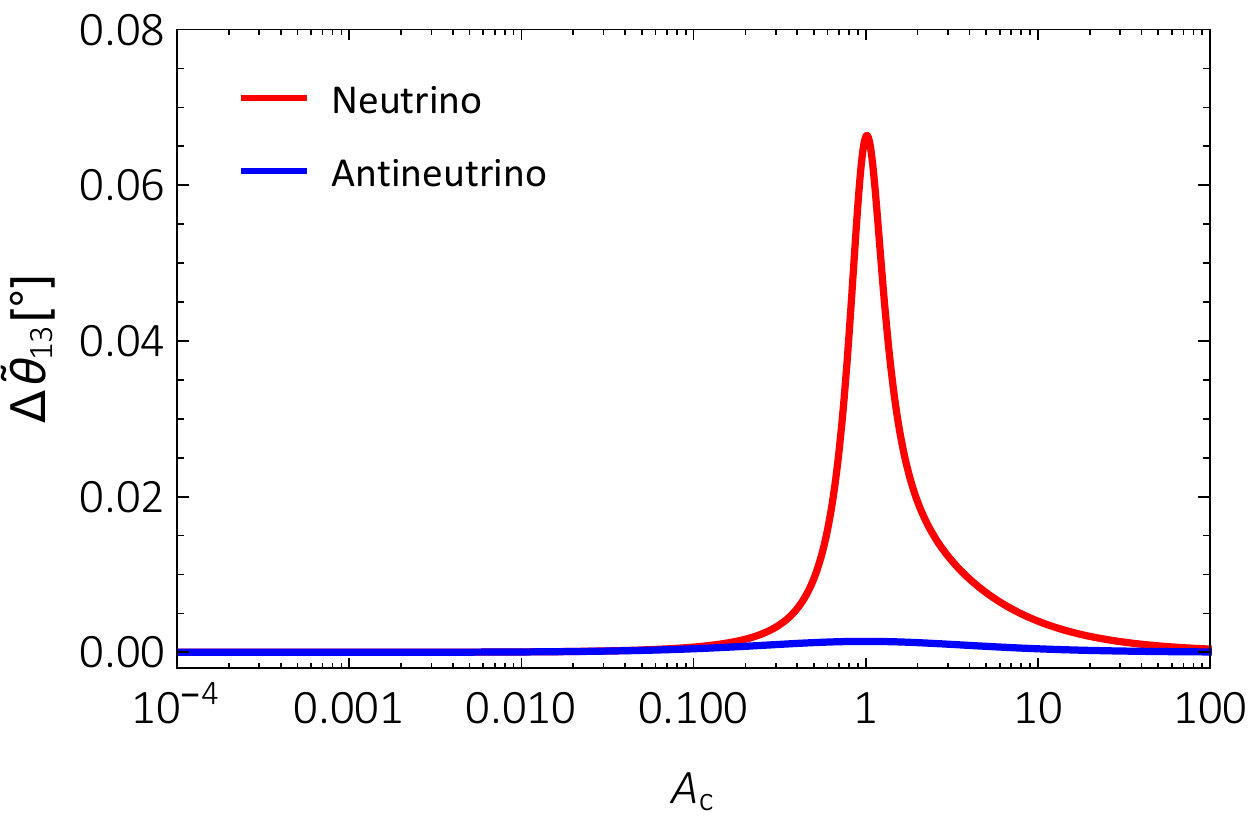}
\vspace{-0.5cm}
\caption{{\it Left panel}: Analytical and numerical solutions to $\widetilde{\theta}^{}_{13}$ in the case of normal neutrino mass ordering, where the best-fit values of neutrino mixing parameters from Ref.~\cite{Esteban:2018azc} have been used in the numerical calculations. The solid curves stand for the analytical solutions, while the dashed ones for the numerical solutions. In addition, the red and orange curves are for neutrinos while the blue and green ones for antineutrinos. {\it Right panel}: The difference $\Delta \widetilde{\theta}^{}_{13} \equiv \widetilde{\theta}_{13}^{} ({\rm analytical}) - \widetilde{\theta}_{13}^{}({\rm numerical})$ between the analytical and numerical results has been plotted as the red curve for neutrinos while the blue curve for antineutrinos.}
	\label{fig:th13} 
\end{figure}

In Fig.~\ref{fig:th13}, the analytical solution to $\widetilde{\theta}^{}_{13}$ in Eq.~(\ref{eq:solth13no}) has been plotted and compared with the exact numerical one. In the left panel, the solid curves stand for the analytical solutions, while the dashed curves for the numerical ones. In addition, the red and orange curves are for neutrinos, while the blue and green ones for antineutrinos. In our numerical calculations, the best-fit values of neutrino mixing parameters $\{\theta^{}_{12} = 33.82^{\circ}, \theta^{}_{13} = 8.61^{\circ}, \theta^{}_{23} = 49.6^{\circ}, \delta= 215^{\circ}\}$ and neutrino mass-squared differences $\{\Delta m^{2}_{21} = 7.39\times 10^{-5}~{\rm eV^2},\Delta m^{2}_{31} = +2.525\times 10^{-3}~{\rm eV^2}\}$ from Ref.~\cite{Esteban:2018azc} have been used. Amazingly, the analytical results are in perfect agreement with the exact numerical ones, indicating that the simple result in Eq.~(\ref{eq:solth13no}) is very accurate over a wide range of $A^{}_{\rm c}$. The difference between analytical and numerical results is shown in the right panel of Fig.~\ref{fig:th13}, where one can see that the largest deviation located around $A^{}_{\rm c}= \cos 2\theta^{}_{13}$ in the neutrino case is no more than $0.08^{\circ}$. The evolution of $\widetilde{\theta}^{}_{13}$ with respect to $A^{}_{\rm c}$ can be well understood with the help of the analytical formula, resembling the main features in the two-flavor neutrino mixing in matter. For antineutrinos, there is no resonance and the effective mixing angle $\widetilde{\theta}^{}_{13}$ will be monotonically decreasing to zero, as it should be suppressed by matter effects. Note that the analytical and numerical results for antineutrinos have been obtained by replacing $A^{}_{\rm c}$ with $-A^{}_{\rm c}$, so the value $A^{}_{\rm c}$ remains to be positive for both neutrinos and antineutrinos, as shown in Fig.~\ref{fig:th13}.

Then, we proceed with the solution to $\widetilde{\theta}^{}_{12}$. As one could expect, it is impossible to get any meaningful results based on the series expansion in Eqs.~(\ref{eq:Del21c})-(\ref{eq:Del32c}). For instance, if we insert Eq.~(\ref{eq:Del21c}) into Eq.~(\ref{eq:rgeDel21}), then it turns out that $\cos 2\widetilde{\theta}^{}_{12} = -1$, given $\cos^2\widetilde{\theta}^{}_{13}$ in Eq.~(\ref{eq:solth13no}). Obviously this solution to $\widetilde{\theta}^{}_{12}$ cannot be correct, as it gives the wrong value of the mixing angle $\theta^{}_{12}$ in vacuum. As we have explained, the series expansion in Eqs.~(\ref{eq:Del21c})-(\ref{eq:Del32c}) is unable to account for the resonance at $A^{}_{\rm c} = \alpha^{}_{\rm c} \cos 2\theta^{}_{12}$, which is however important for $\widetilde{\theta}^{}_{12}$. To this end, we propose a modified version of the effective neutrino mass-squared differences
\begin{eqnarray}
\widetilde{\Delta}^{}_{21} &=& \Delta^{}_{\rm c} \left[ \frac{1}{2} (1 + A^{}_{\rm c} - \widehat{C}^{}_{13}) + \alpha^{}_{\rm c} ({\cal F} - {\cal G}) \right] \; , \label{eq:Del21n} \\
\widetilde{\Delta}_{31} &=& \Delta^{}_{\rm c} \left[ \frac{1}{2} (1 + A^{}_{\rm c} + \widehat{C}^{}_{13}) + \alpha^{}_{\rm c} {\cal F} \right] \; , \label{eq:Del31n} \\
\widetilde{\Delta}_{32} &=& \Delta^{}_{\rm c} \left( \widehat{C}^{}_{13} + \alpha^{}_{\rm c} {\cal G} \right) \; , \label{eq:Del32n}
\end{eqnarray}
where ${\cal F}(A^{}_{\rm c})$ and ${\cal G}(A^{}_{\rm c})$ are two functions of $A^{}_{\rm c}$ that need to be determined. It is worth mentioning that all the terms proportional to $\alpha^{}_{\rm c}$ on the right-hand sides of Eqs.~(\ref{eq:Del21n})-(\ref{eq:Del32n}) are not necessarily regarded as the first-order expansion, since ${\cal F}$ and ${\cal G}$ themselves may depend on $\alpha^{}_{\rm c}$. The reason why we write them in this way is to reproduce the results in Eqs.~(\ref{eq:Del21c})-(\ref{eq:Del32c}) in the limit of large $A^{}_{\rm c}$. On the other hand, we attempt to regularize the effective parameters in the limit of small $A^{}_{\rm c}$ by using the RGEs. Now we explain how to determine these two new functions, which is the central problem to deal with in this work.
\begin{itemize}
\item First, as we have mentioned before, it is safe to implement the series expansion of $\widetilde{m}^2_2 + \widetilde{m}^2_1$ even for small $A^{}_{\rm c}$. Therefore, it is reasonable to demand ${\cal F} + {\cal G} = -\cos 2\theta^{}_{12}$, similar to the situation for Eqs.~(\ref{eq:Del21c})-(\ref{eq:Del32c}). Such a requirement reduces the number of independent new functions from two to one. Moreover, the solution to $\widetilde{\theta}^{}_{13}$ in Eq.~(\ref{eq:solth13no}) agrees excellently with the exact result. In order not to spoil this result, we follow the same procedure leading to Eq.~(\ref{eq:solth13no}) and find that ${\rm d}{\cal F}/{\rm d}A^{}_{\rm c} + {\rm d}{\cal G}/{\rm d}A^{}_{\rm c} = 0$ has to be satisfied. Finally, if $A^{}_{\rm c}$ is set to zero, the effective neutrino mass-squared differences $\widetilde{\Delta}^{}_{ij}$ in Eqs.~(\ref{eq:Del21n})-(\ref{eq:Del32n}) have to recover the neutrino mass-squared differences $\Delta^{}_{ij}$ in vacuum. This gives rise to ${\cal F}(0) = \sin^2 \theta^{}_{12}$ and ${\cal G}(0) = - \cos^2\theta^{}_{12}$, which are the initial conditions necessary for us to determine ${\cal F}$ and ${\cal G}$.

\item Plugging Eq.~(\ref{eq:Del21n}) into Eq.~(\ref{eq:rgeDel21}) and noticing ${\rm d}{\cal G}/{\rm d}A^{}_{\rm c} = -{\rm d}{\cal F}/{\rm d}A^{}_{\rm c}$, we arrive at
\begin{equation}
\cos^2 \widetilde{\theta}^{}_{12} = \frac{2 \widehat{C}^{}_{13} \alpha^{}_{\rm c}}{A^{}_{\rm c} - \widehat{C}^{}_{13} - \cos 2\theta^{}_{13}} \frac{{\rm d}{\cal F}}{{\rm d}A^{}_{\rm c}} \; ,
\label{eq:theta12a}
\end{equation}
where the expression of $\cos^2 \widetilde{\theta}^{}_{13}$ in Eq.~(\ref{eq:solth13no}) has been used. On the other hand, with the help of Eqs.~(\ref{eq:solth13no}), (\ref{eq:Del31n}) and (\ref{eq:Del32n}), one can derive from Eq.~(\ref{eq:rgetheta13}) that
\begin{equation}
\cos^2 \widetilde{\theta}^{}_{12} = \frac{(1 + A^{}_{\rm c} + \widehat{C}^{}_{13}) (\cos 2\theta^{}_{12} + {\cal F}) \alpha^{}_{\rm c}}{\widehat{C}^{}_{13} \left[2(\cos 2\theta^{}_{12} + 2{\cal F})\alpha^{}_{\rm c} + (1 + A^{}_{\rm c} - \widehat{C}^{}_{13})\right]} \; ,
\label{eq:theta12b}
\end{equation}
where the term proportional to $\alpha^2_{\rm c}$ in the numerator has been neglected. By identifying the right-hand side of Eq.~(\ref{eq:theta12a}) with that of Eq.~(\ref{eq:theta12b}), we can establish the differential equation
\begin{equation}
\frac{{\rm d}{\cal F}}{{\rm d}A^{}_{\rm c}} = \frac{(A^{}_{\rm c} - \widehat{C}^{}_{13} - 1) (\cos 2\theta^{}_{12} + {\cal F}) \cos^2 \theta^{}_{13} } {\widehat{C}^2_{13} \left[ 2(\cos 2\theta^{}_{12} + 2{\cal F}) \alpha^{}_{\rm c} + (1 + A^{}_{\rm c} - \widehat{C}^{}_{13})\right]} \;.
\label{eq:dFdA}
\end{equation}
Usually it is difficult to solve Eq.~(\ref{eq:dFdA}) in the most general case. For simplicity, we look for the solution in the limit of $A^{}_{\rm c} \to 0$. In this limit, one can easily check that $(A^{}_{\rm c} - \widehat{C}^{}_{13} - 1)/\widehat{C}^2_{13} \approx -2$ and $(1 + A^{}_{\rm c} - \widehat{C}^{}_{13})/2 \approx A^{}_{\rm c} \cos^2 \theta^{}_{13}$. As a consequence, Eq.~(\ref{eq:dFdA}) will be considerably simplified to
\begin{eqnarray}
\frac{{\rm d}{\cal F}}{{\rm d}A^{}_{\rm c}} = - \frac{(\cos 2\theta^{}_{12} + {\cal F}) \cos^2 \theta^{}_{13}}{(\cos 2\theta^{}_{12} + 2{\cal F})\alpha^{}_{\rm c} + A^{}_{\rm c}} \; ,
\label{eq:dFdAsim}
\end{eqnarray}
where the terms proportional to $A^{}_{\rm c} \sin^2 \theta^{}_{13}$ have been safely ignored. The exact solution to Eq.~(\ref{eq:dFdAsim}) can then be found if $\cos^2 \theta^{}_{13} \approx 1$ is further assumed, i.e.,
\begin{eqnarray}
{\cal F}(A^{}_{\rm c}) = \frac{1}{2\alpha^{}_{\rm c}}\left[\sqrt{(A^{}_{\rm c} - \alpha^{}_{\rm c} \cos 2\theta^{}_{12})^2 + \alpha^2_{\rm c} \sin^2 2\theta^{}_{12}} - (A^{}_{\rm c} + \alpha^{}_{\rm c} \cos 2\theta^{}_{12})\right] \; .
\label{eq:solF}
\end{eqnarray}
It should be noticed that although Eq.~(\ref{eq:solF}) is simple, it correctly reproduces ${\cal F} \to \sin^2 \theta^{}_{12}$ in the limit $A^{}_{\rm c} \to 0$ and ${\cal F} \to -\cos 2\theta^{}_{12}$ in the limit $A^{}_{\rm c} \to \infty$. Without the assumption of $\cos^2 \theta^{}_{13} \approx 1$, one can still analytically solve the differential equation in Eq.~(\ref{eq:dFdAsim}), but the final solution will be more complicated and less useful. The other function is then given by ${\cal G}(A^{}_{\rm c}) = -\cos 2\theta^{}_{12} - {\cal F}(A^{}_{\rm c})$, implying ${\cal G}(A^{}_{\rm c})|^{}_{A^{}_{\rm c} \to 0} \to -\cos^2 \theta^{}_{12}$ and ${\cal G}(A^{}_{\rm c})|^{}_{A^{}_{\rm c} \to \infty} \to 0$.
\end{itemize}
\begin{figure}[!t]
	\centering
	\includegraphics[width=0.49\textwidth]{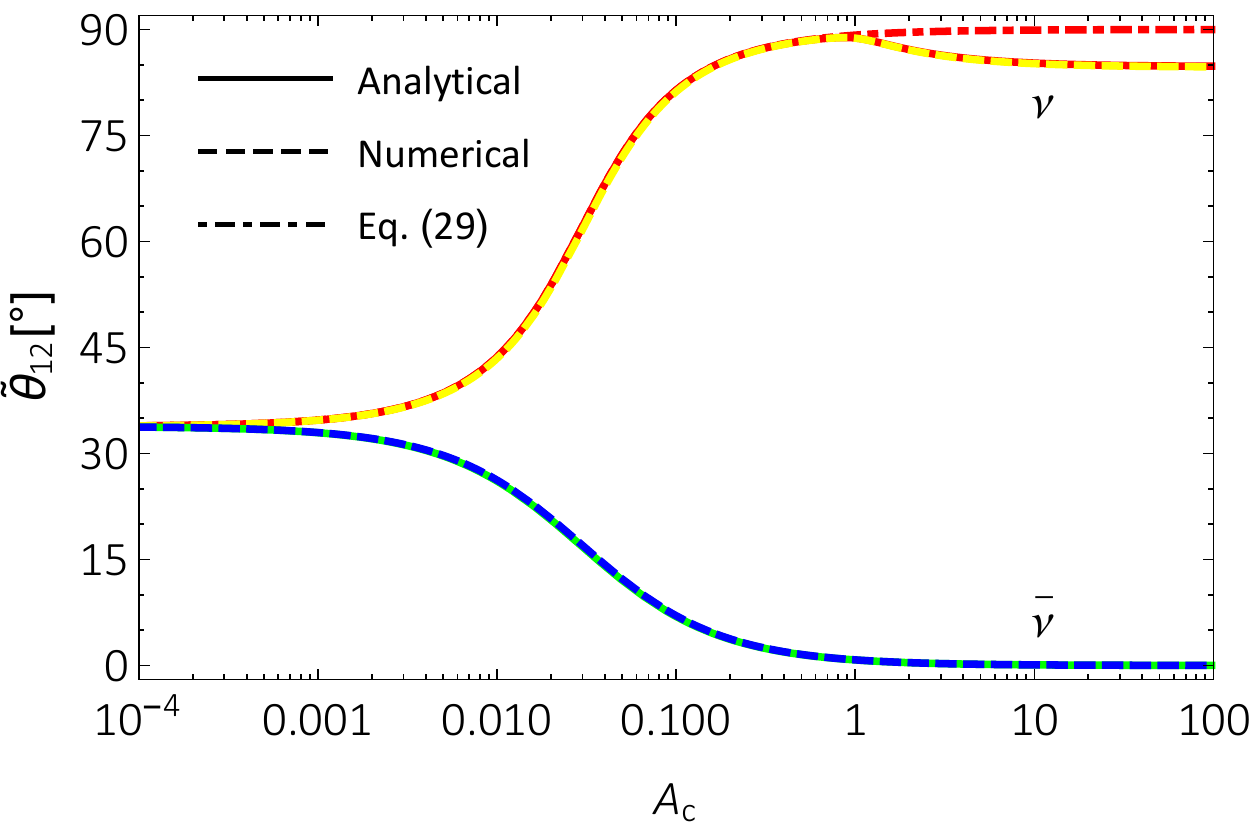}
	\includegraphics[width=0.50\textwidth]{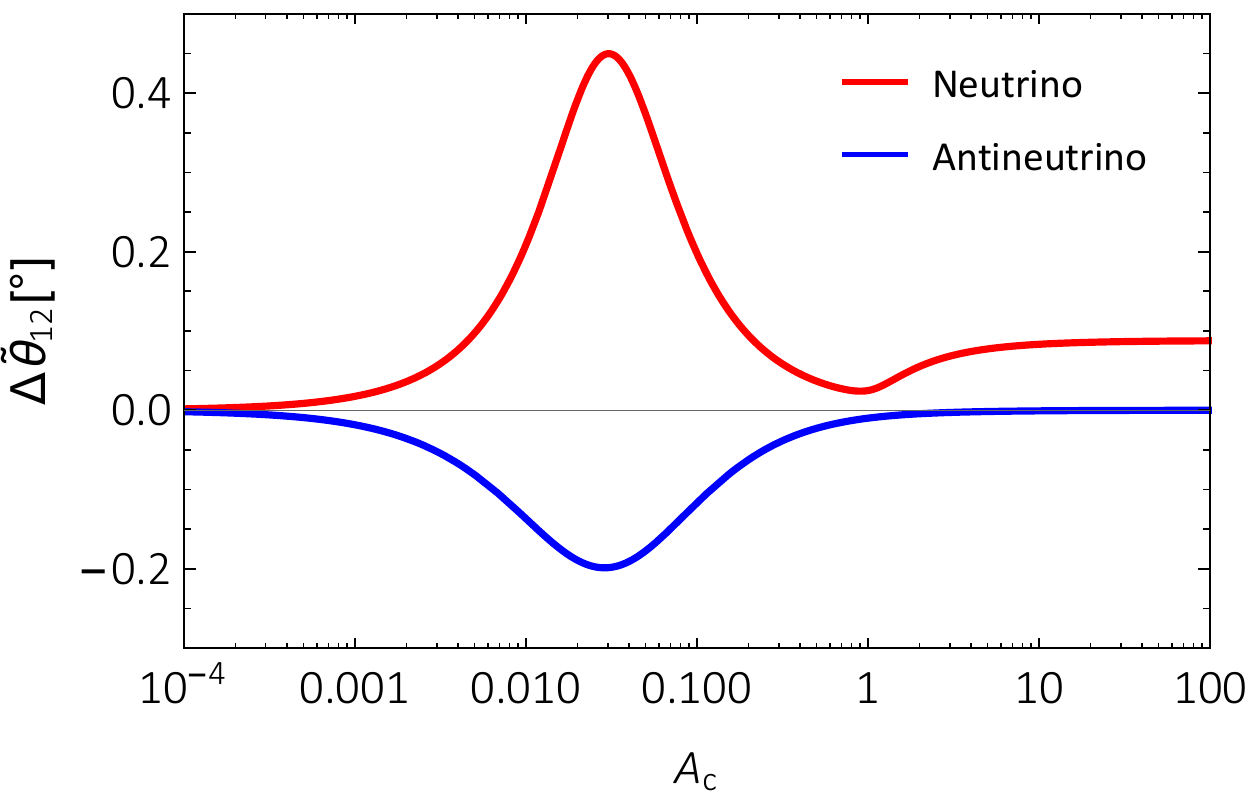}
	\vspace{-0.5cm}
	\caption{{\it Left panel}: Analytical and numerical solutions to $\widetilde{\theta}^{}_{12}$ in the case of normal neutrino mass ordering, where the input parameters are the same as in Fig.~\ref{fig:th13}. For the analytical solutions, the approximate result in Eq.~(\ref{eq:solth12app}) has been plotted as the red dotted-dashed curve, while the full result in Eq.~(\ref{eq:solth12}) as the red solid curve. {\it Right panel}: The difference $\Delta \widetilde{\theta}^{}_{12}$ between the full result in Eq.~(\ref{eq:solth12}) and the numerical result is plotted as the red curve for neutrinos while the blue one for antineutrinos.}
	\label{fig:th12} 
\end{figure}

With the function ${\cal F}(A^{}_{\rm c})$ in Eq.~(\ref{eq:solF}), we can substitute it into the right-hand side of Eq.~(\ref{eq:dFdAsim}) and then insert the expression of ${\rm d}{\cal F}/{\rm d}A^{}_{\rm c}$ into Eq.~(\ref{eq:theta12a}), leading to the ultimate solution of $\widetilde{\theta}^{}_{12}$
\begin{equation}
\cos^2 \widetilde{\theta}^{}_{12} =  \frac{1}{2} \left(1 - \frac{A^{}_* - \cos 2\theta^{}_{12}}{\widehat{C}^{}_{12}}\right) \frac{2\widehat{C}^{}_{13} \cos^2 \theta^{}_{13}}{\widehat{C}^{}_{13} - A^{}_{\rm c} + \cos 2\theta^{}_{13}} \; ,
\label{eq:solth12}
\end{equation}
where $A^{}_* \equiv A^{}_{\rm c}/\alpha^{}_{\rm c} = a/\Delta^{}_{21}$ and $\widehat{C}^{}_{12} \equiv \sqrt{1 - 2A^{}_* \cos 2\theta^{}_{12} + A^2_*} = \sqrt{(A^{}_* - \cos 2\theta^{}_{12})^2 + \sin^2 2\theta^{}_{12}}$ have been introduced. In Eq.~(\ref{eq:solth12}), the last term on the right-hand side can also be written as $\cos^2 \theta^{}_{13}/\cos^2 \widetilde{\theta}^{}_{13}$, which deserves further discussions. As we have seen from Fig.~\ref{fig:th13}, $\widetilde{\theta}^{}_{13}$ in the neutrino case increases very slowly until the resonance at $A^{}_{\rm c} = \cos 2\theta^{}_{13}$ is reached, then it will be resonantly enhanced to $90^\circ$. Since the width of the resonance is extremely narrow, as it is characterized by the smallest mixing angle $\theta^{}_{13} \approx 8^\circ$, one can simply take $\cos^2 \theta^{}_{13}/\cos^2 \widetilde{\theta}^{}_{13} \approx 1$ for $A^{}_{\rm c} < \cos 2\theta^{}_{13}$ and then get
\begin{equation}
\cos^2 \widetilde{\theta}^{}_{12} =  \frac{1}{2} \left(1 - \frac{A^{}_* - \cos 2\theta^{}_{12}}{\widehat{C}^{}_{12}}\right) \; .
\label{eq:solth12app}
\end{equation}
Making a comparison between Eq.~(\ref{eq:solth12app}) and Eq.~(\ref{eq:solth13no}), we realize that $\widetilde{\theta}^{}_{12}$ can also be described by the effective mixing angle in matter in the limit of two-flavor mixing with $\theta^{}_{12}$ being the mixing angle and $\Delta^{}_{21}$ being the relevant mass-squared difference in vacuum~\cite{Akhmedov2004}. However, the correction factor $\cos^2 \theta^{}_{13}/\cos^2 \widetilde{\theta}^{}_{13}$ becomes significant when approaching the resonance at $A^{}_{\rm c} = \cos 2\theta^{}_{13}$.

In Fig.~\ref{fig:th12}, the analytical and numerical solutions to $\widetilde{\theta}^{}_{12}$ have been shown and compared with each other. In the left panel, the approximate analytical result in Eq.~(\ref{eq:solth12app}) and the full analytical result in Eq.~(\ref{eq:solth12}) are plotted as the red dotted-dashed and solid curve, respectively. In the right panel, the difference between the full analytical result and the numerical result has been shown, where the largest deviation appearing in the resonance region at $A^{}_{\rm c} = \alpha^{}_{\rm c} \cos 2\theta^{}_{12}$ is about $\Delta \widetilde{\theta}^{}_{12} \approx 0.5^\circ$ for neutrinos and $\Delta \widetilde{\theta}^{}_{12} \approx 0.2^\circ$ for antineutrinos. As has been pointed out before, the difference between Eq.~(\ref{eq:solth12}) and Eq.~(\ref{eq:solth12app}) is the inclusion of the correction factor $\cos^2 \theta^{}_{13}/\cos^2 \widetilde{\theta}^{}_{13}$ in the former equation, which changes the asymptotic behavior of $\widetilde{\theta}^{}_{12}$ in the limit $A^{}_{\rm c} \to \infty$. To be explicit, Eq.~(\ref{eq:solth12app}) implies $\cos^2 \widetilde{\theta}^{}_{12} \to 0$ or equivalently $\widetilde{\theta}^{}_{12} \to 90^\circ$ for $A^{}_{\rm c} \to \infty$. However, Eq.~(\ref{eq:solth12}) gives rise to $\cos^2 \widetilde{\theta}^{}_{12} \to \alpha^2_{\rm c} \csc^2 \theta^{}_{13} \sin^2 \theta^{}_{12} \cos^2 \theta^{}_{12}$ or equivalently $\widetilde{\theta}^{}_{12} \to 84.6^\circ$ in the limit $A^{}_{\rm c} \to \infty$. 

In addition, $\widetilde{\theta}^{}_{12}$ in the antineutrino case has also been presented in Fig.~\ref{fig:th12}, where the analytical results in both left and right panels have been obtained from Eq.~(\ref{eq:solth12app}) by setting $A^{}_* \to -A^{}_*$. Since there is no resonance for antineutrino oscillations in matter in the NO case, the analytical solution derived from Eq.~(\ref{eq:solth12app}) works pretty well. As one can observe from the left panel of Fig.~\ref{fig:th12}, the matter effects tend to suppress the effective mixing angle $\widetilde{\theta}^{}_{12}$. This is expected in general if the resonance is absent.

Now that the functions ${\cal F}(A^{}_{\rm c})$ and ${\cal G}(A^{}_{\rm c})$ have been determined, it is straightforward to rewrite the expressions of effective neutrino mass-squared differences as
\begin{eqnarray}
\widetilde{\Delta}^{}_{21} &=& \Delta^{}_{\rm c} \left[\frac{1}{2}(1 + A^{}_{\rm c} - \widehat{C}^{}_{13}) + (\widehat{C}^{}_{12} - A^{}_*) \alpha^{}_{\rm c}\right] \; , \label{eq:Del21rev} \\
\widetilde{\Delta}^{}_{31} &=& \Delta^{}_{\rm c} \left[\frac{1}{2}(1 + A^{}_{\rm c} + \widehat{C}^{}_{13}) + \frac{1}{2} (\widehat{C}^{}_{12} - A^{}_* - \cos 2\theta_{12}) \alpha^{}_{\rm c} \right] \; , \label{eq:Del31rev} \\ \widetilde{\Delta}^{}_{32} &=& \Delta^{}_{\rm c} \left[\widehat{C}^{}_{13} + \frac{1}{2} (A^{}_* - \widehat{C}^{}_{12} - \cos 2\theta_{12}) \alpha^{}_{\rm c} \right] \; . \label{eq:Del32rev}
\end{eqnarray}
\begin{figure}[t!]
	\centering
	\includegraphics[width=0.49\textwidth]{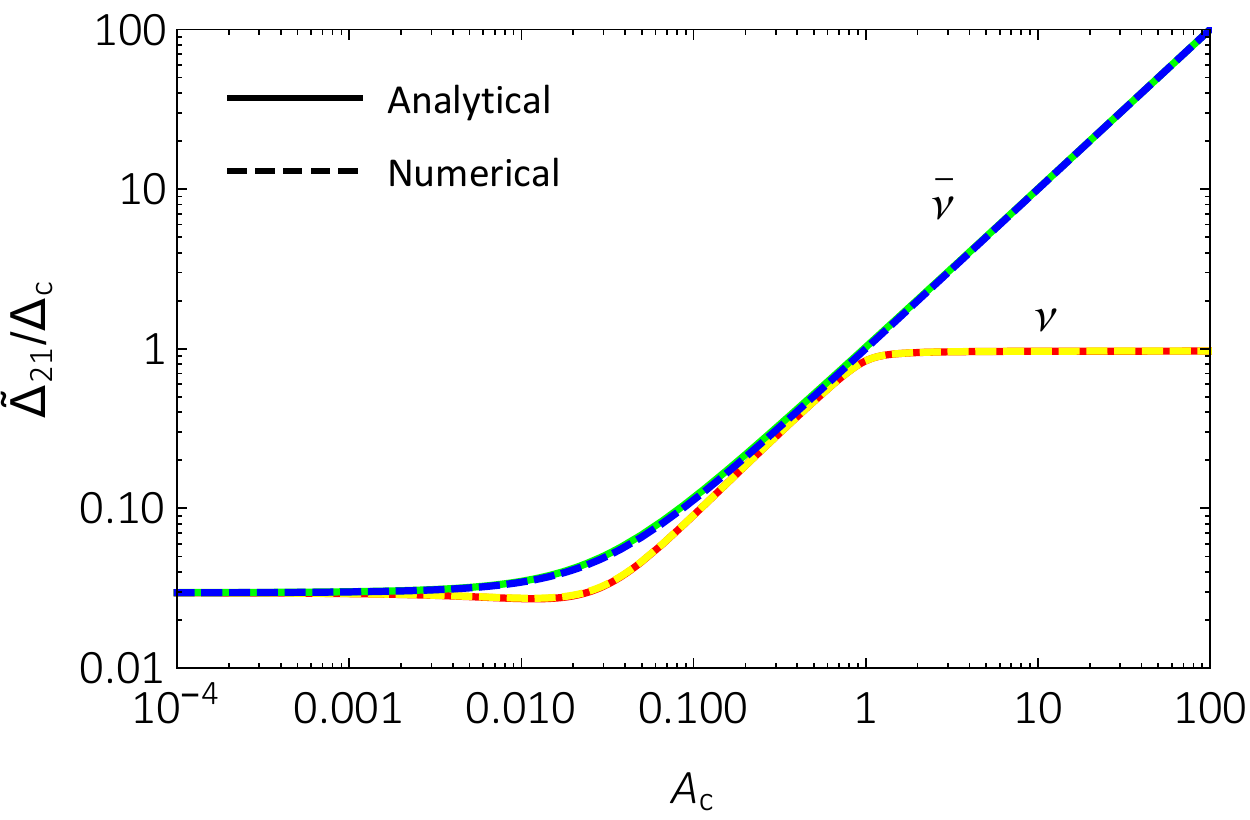}
	\includegraphics[width=0.49\textwidth]{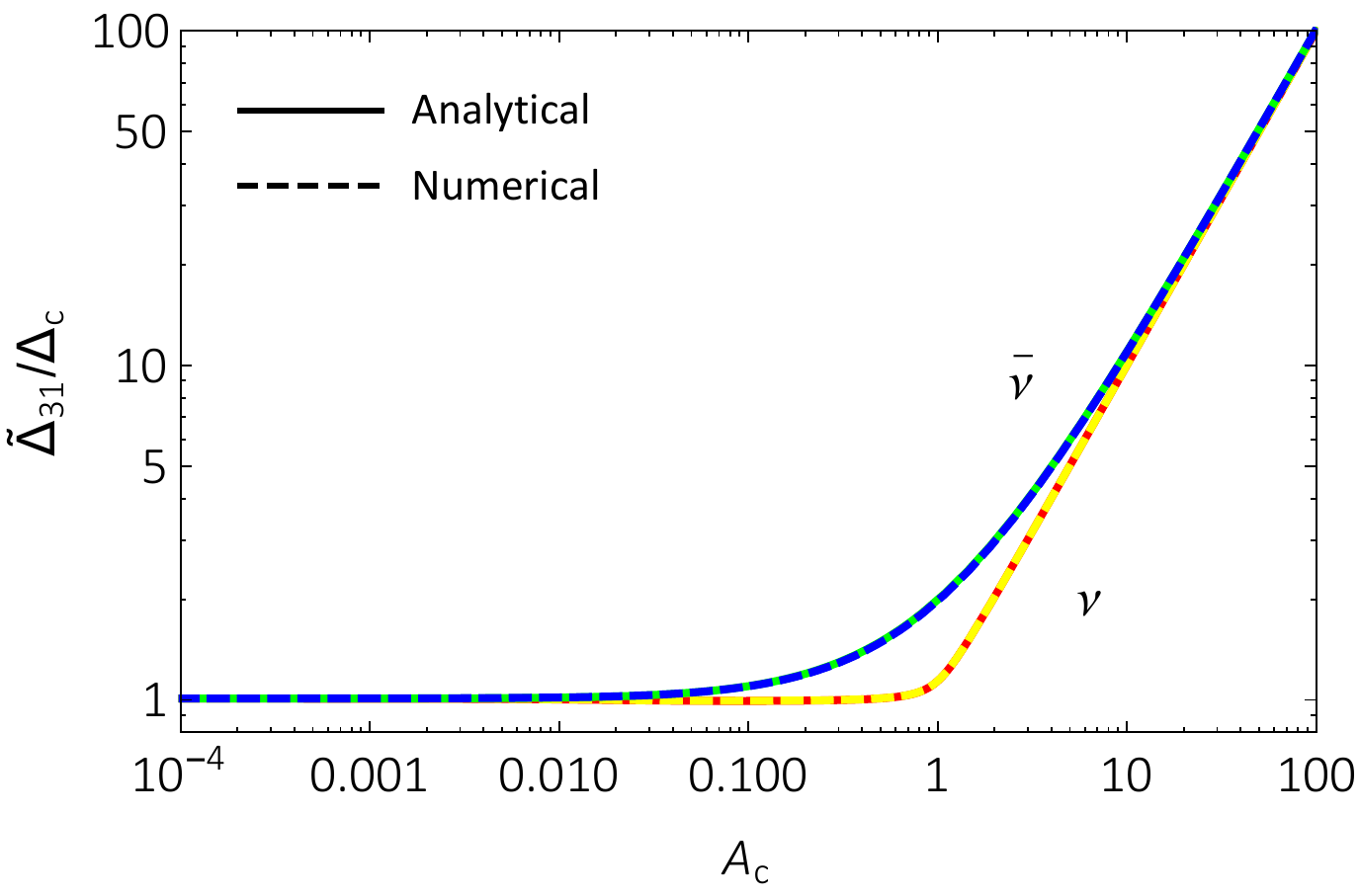}
	\vspace{-0.2cm}
	\caption{Analytical and numerical solutions to $\widetilde{\Delta}^{}_{21}/\Delta^{}_{\rm c}$ ({\it left panel}) and $\widetilde{\Delta}^{}_{31}/\Delta^{}_{\rm c}$ ({\it right panel}) in the case of normal neutrino mass ordering, where the input parameters and the conventions for the curves are the same as in the previous figures.}
	\label{fig:Deltaij} 
\end{figure}
For completeness, two independent effective neutrino mass-squared differences $\widetilde{\Delta}^{}_{21}/\Delta^{}_{\rm c}$ and $\widetilde{\Delta}^{}_{31}/\Delta^{}_{\rm c}$ are shown against the normalized matter parameter $A^{}_{\rm c} = a/\Delta^{}_{\rm c}$ in Fig.~\ref{fig:Deltaij}, where the analytical results in Eqs.~(\ref{eq:Del21rev}) and (\ref{eq:Del31rev}) have been compared with their exact values. An excellent agreement between analytical and numerical results can be observed for the whole range of $10^{-4} \lesssim A^{}_{\rm c} \lesssim 10^2$. The evolution of $\widetilde{\Delta}^{}_{21}/\Delta^{}_{\rm c}$ and $\widetilde{\Delta}^{}_{31}/\Delta^{}_{\rm c}$ can be well understood from their RGEs in Eq.~(\ref{eq:rgeDel21}) and Eq.~(\ref{eq:rgeDel31}), respectively, given the solutions of $\widetilde{\theta}^{}_{13}$ and $\widetilde{\theta}^{}_{12}$. For neutrino oscillations in matter, we have ${\rm d}(\widetilde{\Delta}^{}_{21}/\Delta^{}_{\rm c})/{\rm d}A^{}_{\rm c} = - \cos^2 \widetilde{\theta}^{}_{13} \cos 2\widetilde{\theta}^{}_{12}$, so $\widetilde{\Delta}^{}_{21}/\Delta^{}_{\rm c}$ first decreases slowly until the resonance at $A^{}_{\rm c} = \alpha^{}_{\rm c} \cos 2\theta^{}_{12}$ when $\cos 2\widetilde{\theta}^{}_{12}$ changes its sign. Later on, when another resonance $A^{}_{\rm c} = \cos 2\theta^{}_{13}$ is reached, $\cos^2 \widetilde{\theta}^{}_{13}$ approaches zero and thus $\widetilde{\Delta}^{}_{21}/\Delta^{}_{\rm c}$ arrives at its maximum and becomes stable at this point. In a similar way, one can investigate the evolution of $\widetilde{\Delta}^{}_{31}/\Delta^{}_{\rm c}$ with respect to $A^{}_{\rm c}$.

For antineutrino oscillations in matter, in addition to the absence of resonances, the overall sign change on the left-hand sides of the RGEs should be noticed when $A^{}_{\rm c} \to -A^{}_{\rm c}$ is set. It can be observed from Fig.~\ref{fig:th13} and Fig.~\ref{fig:th12} that the evolution of $\widetilde{\theta}^{}_{13}$ and $\widetilde{\theta}^{}_{12}$ is very simple, namely, decreasing monotonically for increasing $A^{}_{\rm c}$. For $\widetilde{\Delta}^{}_{21}/\Delta^{}_{\rm c}$ and $\widetilde{\Delta}^{}_{31}/\Delta^{}_{\rm c}$ in Fig.~\ref{fig:Deltaij}, they turn out to be linearly proportional to $A^{}_{\rm c}$ after the corresponding resonance is passed.

\subsection{$\widetilde{\theta}^{}_{23}$ and $\widetilde{\delta}$}

Since the analytical solutions to two neutrino effective mixing angles $\{\widetilde{\theta}^{}_{12}, \widetilde{\theta}^{}_{13}\}$ and three effective neutrino mass-squared differences $\{\widetilde{\Delta}^{}_{21}, \widetilde{\Delta}^{}_{31}, \widetilde{\Delta}^{}_{32}\}$ have been found, it is time to solve the remaining two parameters $\widetilde{\theta}^{}_{23}$ and $\widetilde{\delta}$. However, if we simply substitute the analytical expressions into the RGEs of $\widetilde{\theta}^{}_{23}$ and $\widetilde{\delta}$ in Eqs.~(\ref{eq:rgetheta23}) and (\ref{eq:rgedelta}), it will be too complicated to find any analytical and useful results. For this reason, we have to make some reasonable approximations.

From Eqs.~(\ref{eq:rgetheta23}) and (\ref{eq:rgedelta}), one can observe that the evolution of $\widetilde{\theta}^{}_{23}$ and $\widetilde{\delta}$ will be suppressed by both $\sin \theta^{}_{13}$ and $\Delta^{}_{21}/\Delta^{}_{31}$, at least in the region of small $A^{}_{\rm c}$. Therefore, $\widetilde{\theta}^{}_{23} \approx \theta^{}_{23}$ and $\widetilde{\delta} \approx \delta$ can serve as the zeroth-order solutions. It is worthwhile to emphasize that such approximations hardly affect the previous results for the other effective parameters, since their RGEs are independent of $\widetilde{\theta}^{}_{23}$ and $\widetilde{\delta}$. Therefore, it is reasonable to take $\widetilde{\delta} = \delta$ on the right-hand side of Eq.~(\ref{eq:rgetheta23}), namely,
\begin{equation}
\frac{{\rm d}\widetilde{\theta}^{}_{23}}{{\rm d}a} = \frac{\sin 2\widetilde{\theta}^{}_{12} \sin \widetilde{\theta}^{}_{13} \widetilde{\Delta}^{}_{21}}{2 \widetilde{\Delta}^{}_{31} \widetilde{\Delta}^{}_{32}} \cos\delta \; .
\label{eq:deltath23}
\end{equation}
Once $A^{}_{\rm c}$ becomes larger, $\widetilde{\Delta}^{}_{21}$ and $\sin^2 \widetilde{\theta}^{}_{13}$ start to increase. In this case, the approximate result $\widetilde{\theta}^{}_{23} \approx \theta^{}_{23}$ is no longer valid. To explore the evolution of $\widetilde{\theta}^{}_{23}$ in the region of large $A^{}_{\rm c}$, we expand the analytical solutions to $\sin 2\widetilde{\theta}^{}_{12}$, $\sin \widetilde{\theta}^{}_{13}$ and $\widetilde{\Delta}^{}_{ij}$ (for $ij = 21, 31, 32$) on the right-hand side of Eq.~(\ref{eq:deltath23}) as a series of $1/A^{}_{\rm c}$ and retain only the leading-order terms. As a consequence, Eq.~(\ref{eq:deltath23}) turns out to be
\begin{figure}[t!]
	\centering
	\includegraphics[width=0.49\textwidth]{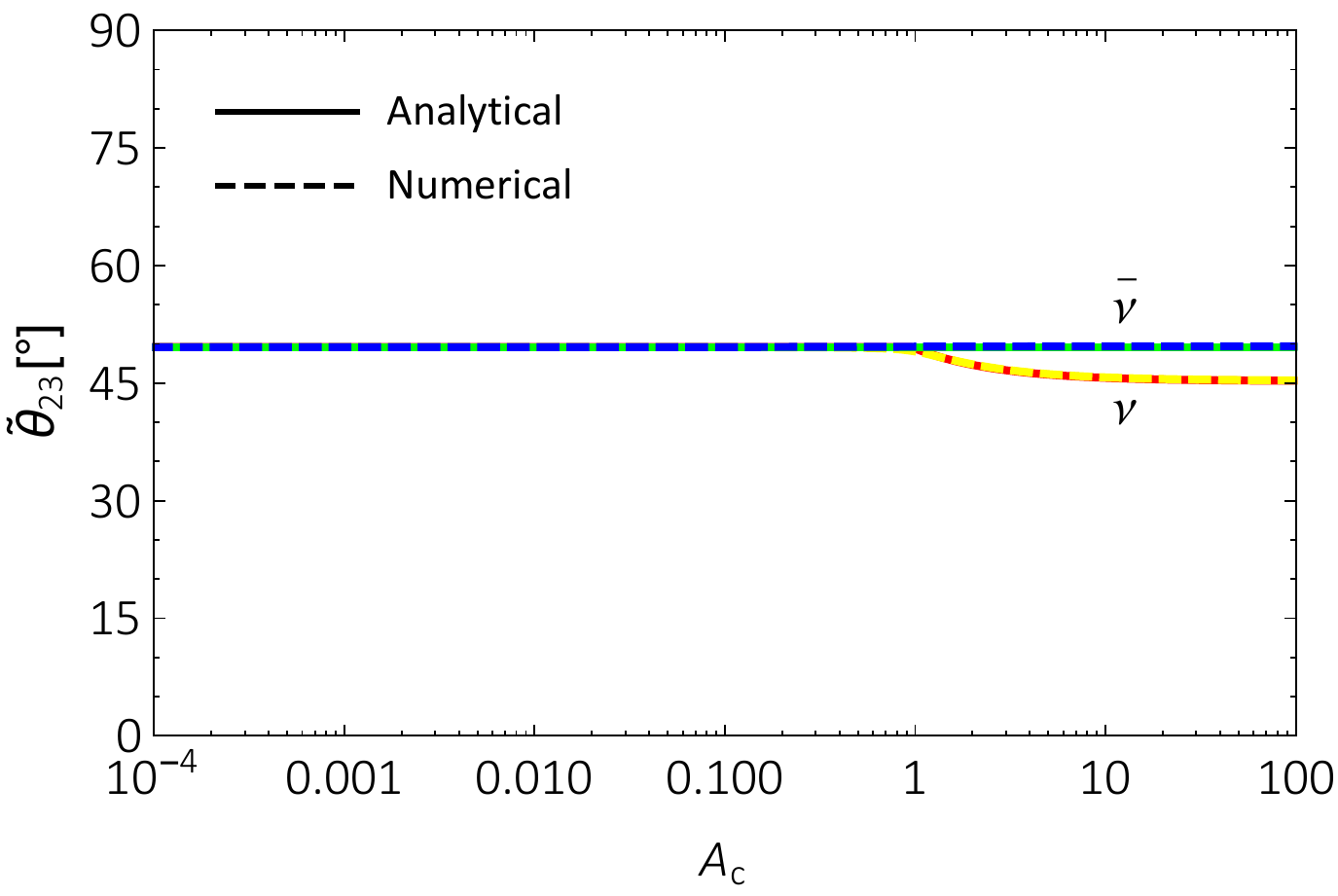}
	\includegraphics[width=0.495\textwidth]{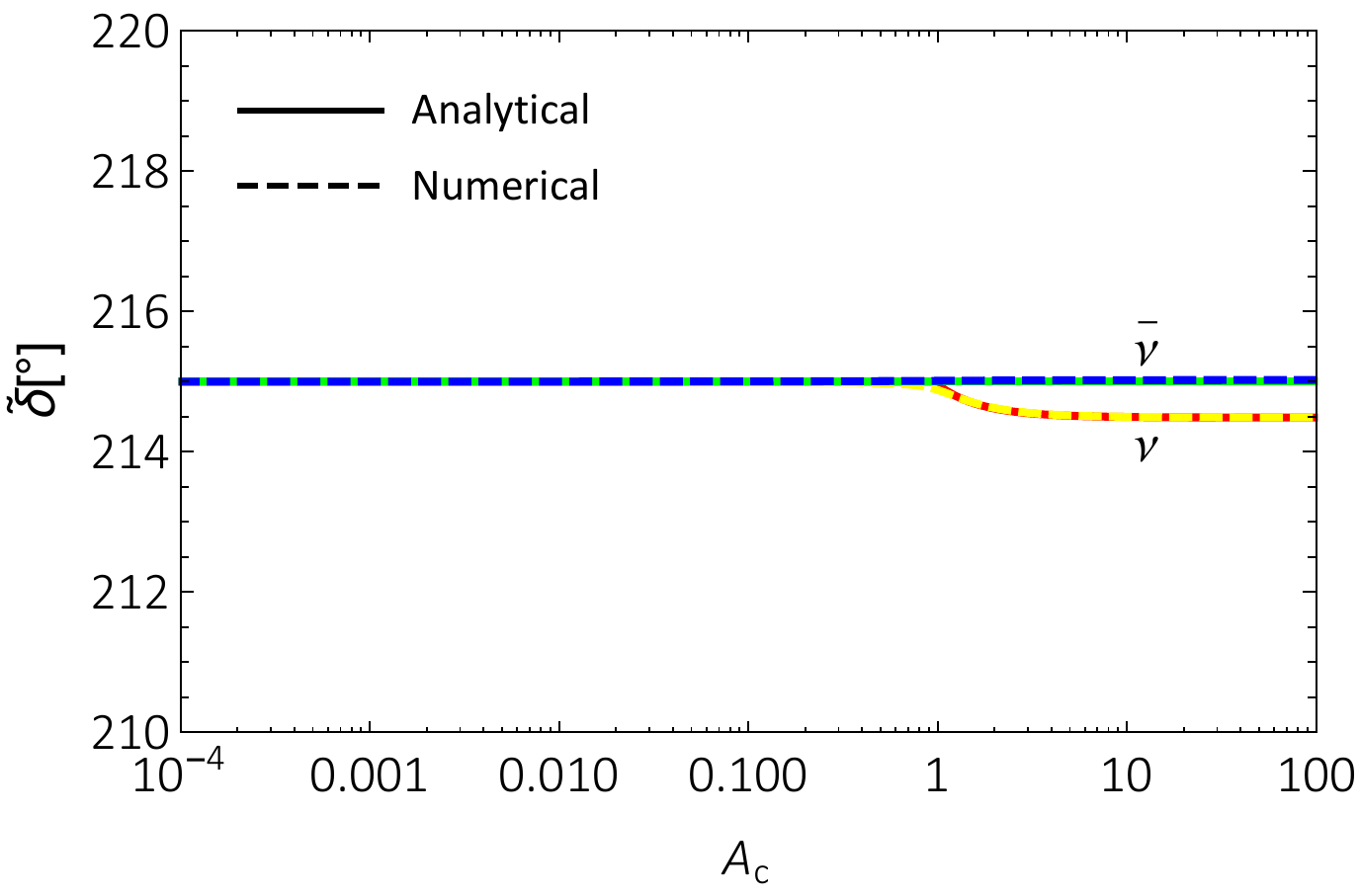}
	\vspace{-0.2cm}
	\caption{Analytical and numerical solutions to $\widetilde{\theta}^{}_{23}$ ({\it left panel}) and $\widetilde{\delta}$ ({\it right panel}), where the input parameters and conventions for the curves are the same as in the previous figures.}
	\label{fig:th23} 
\end{figure}
\begin{equation}
\frac{{\rm d}\widetilde{\theta}^{}_{23}}{{\rm d}A^{}_{\rm c}} = \frac{\alpha^{}_{\rm c} \sin 2\theta^{}_{12} (\cos^2 \theta^{}_{13} - \alpha^{}_{\rm c} \cos 2\theta^{}_{12})}{2 A^2_{\rm c} \sin\theta^{}_{13}} \cos \delta \; .
\label{eq:deltath23f}
\end{equation}
It is easy to solve Eq.~(\ref{eq:deltath23f}) with the initial condition $\widetilde{\theta}^{}_{23}(A^{}_{\rm c})|^{}_{A^{}_{\rm c} = \cos 2\theta^{}_{13}} = \theta^{}_{23}$. Therefore, we obtain the final solution $\widetilde{\theta}^{}_{23} = \theta^{}_{23}$ for $A^{}_{\rm c} \leq \cos 2\theta^{}_{13}$; and
\begin{equation}
\widetilde{\theta}^{}_{23} =
\theta^{}_{23} + \frac{\alpha^{}_{\rm c} \sin2\theta^{}_{12} (\cos^2\theta^{}_{13} - \alpha^{}_{\rm c} \cos2\theta^{}_{12})} {2A^{}_{\rm c} \cos2\theta^{}_{13} \sin\theta^{}_{13}} (A^{}_{\rm c} - \cos2\theta^{}_{13}) \cos \delta \; ,
\label{eq:solth23}
\end{equation}
for $A^{}_{\rm c} > \cos 2\theta^{}_{13}$. In the case of antineutrino oscillations, $\sin \widetilde{\theta}^{}_{13}$ is always small such that $\widetilde{\theta}^{}_{23} = \theta^{}_{23}$ holds excellently for the whole range of $A^{}_{\rm c}$. After getting the analytical result for $\widetilde{\theta}^{}_{23}$, we can calculate $\widetilde{\delta}$ by using the well-known Toshev relation~\cite{Toshev:1991ku}
\begin{equation}
\sin \widetilde{\delta} = \frac{\sin 2\theta^{}_{23}}{\sin 2\widetilde{\theta}^{}_{23}} \sin \delta \; ,
\label{eq:soldelta}
\end{equation}
which is applicable for both neutrinos and antineutrinos. The analytical and numerical solutions to $\widetilde{\theta}^{}_{23}$ and $\widetilde{\delta}$ are given in Fig.~\ref{fig:th23} in the left and right panel, respectively, where one can observe an excellent agreement. We have also checked that the differences between analytical and numerical results are on the level of ${\cal O}(10^{-2})$ degrees. 

\begin{figure}[t!]
\centering
\includegraphics[width=0.49\textwidth]{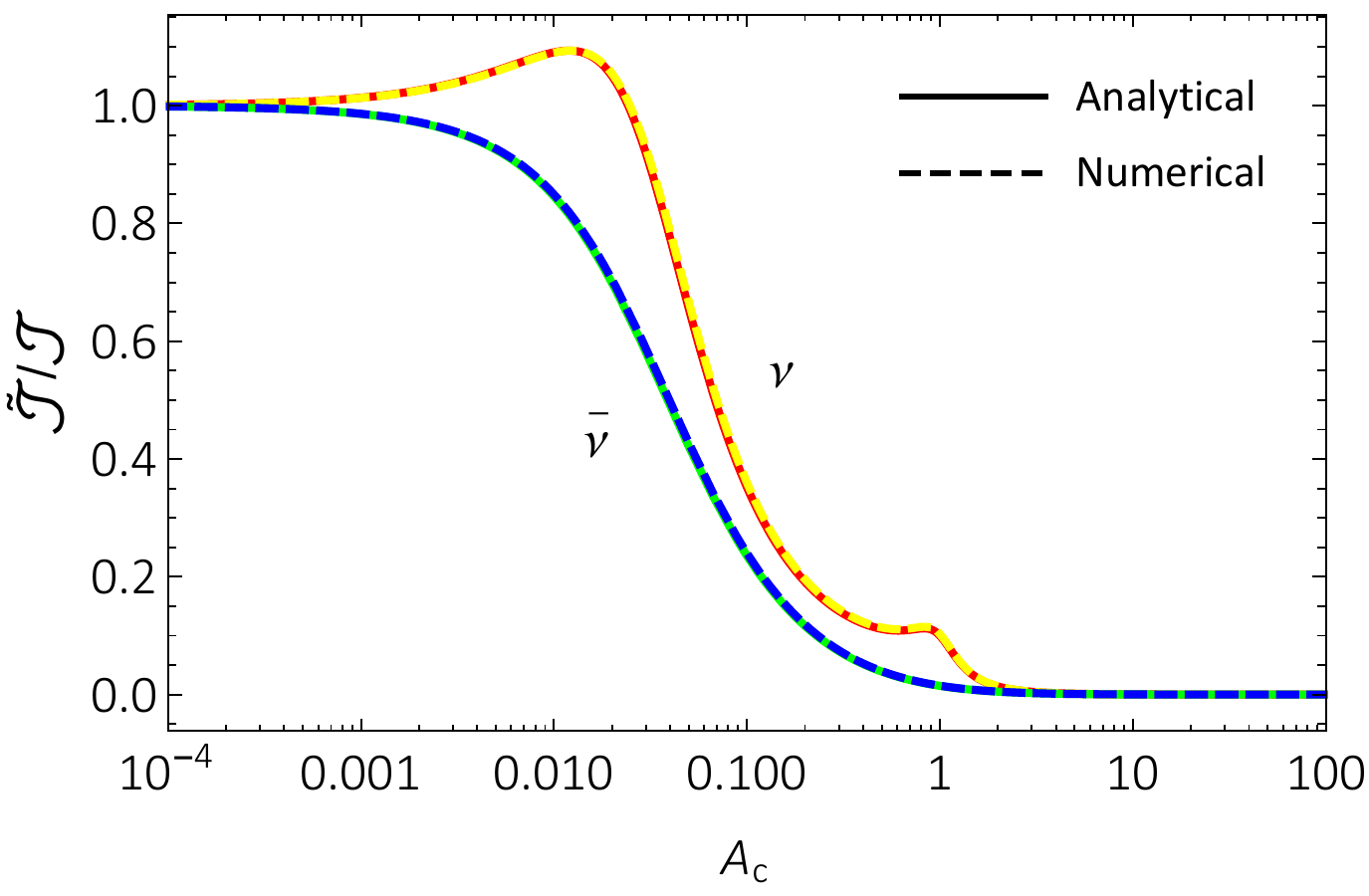}
\vspace{-0.2cm}
\caption{Analytical and numerical solutions to $\widetilde{\cal J}/{\cal J}$ in the case of normal neutrino mass ordering,  where the input parameters and conventions for the curves are the same as in the previous figures.}
\label{fig:J} 
\end{figure}

As an intriguing byproduct, the approximate formula of the Jarlskog invariant for CP violation in the lepton sector~\cite{Jarlskog:1985ht, Wu:1985ea} has been found to be very simple. In the standard parametrization of the effective mixing matrix, the Jarlskog invariant $\widetilde{\cal J}$ in matter can be written as
\begin{equation}
\widetilde{\mathcal{J}} = \sin \widetilde{\theta}^{}_{12} \cos \widetilde{\theta}^{}_{12} \sin \widetilde{\theta}^{}_{13} \cos^2 \widetilde{\theta}^{}_{13} \sin \widetilde{\theta}^{}_{23} \cos \widetilde{\theta}^{}_{23} \sin \widetilde{\delta} \; .
\label{eq:jarls}
\end{equation}
The Jarlskog invariant ${\cal J}$ in vacuum can be obtained by replacing the effective mixing parameters in Eq.~(\ref{eq:jarls}) with their counterparts in vacuum. Given the Toshev relation in Eq.~(\ref{eq:soldelta}) and the analytical solutions to the effective mixing parameters, the ratio $\widetilde{\cal J}/{\cal J}$ is found to be
\begin{equation}
\frac{\widetilde{\mathcal{J}}}{\mathcal{J}} = \frac{\sin \widetilde{\theta}^{}_{12} \cos \widetilde{\theta}^{}_{12} \sin \widetilde{\theta}^{}_{13} \cos^2 \widetilde{\theta}^{}_{13}}{\sin \theta^{}_{12} \cos \theta^{}_{12} \sin \theta^{}_{13} \cos^2 \theta^{}_{13}} \approx \frac{1}{\widehat{C}^{}_{12} \widehat{C}^{}_{13}} \; ,
\label{eq:jratio}
\end{equation}
where the Toshev relation has been used in the derivation of the first identity. For antineutrinos, one can make the transformation $A^{}_{\rm c} \to -A^{}_{\rm c}$ in $\widehat{C}^{}_{13}$ and $A^{}_* \to -A^{}_*$ in $\widehat{C}^{}_{12}$. If the exact Naumov relation ${\cal J} \Delta^{}_{21} \Delta^{}_{31} \Delta^{}_{32} = \widetilde{\cal J} \widetilde{\Delta}^{}_{21} \widetilde{\Delta}^{}_{31} \widetilde{\Delta}^{}_{32}$ is implemented~\cite{Naumov:1991ju, Harrison:1999df, Xing:2000gg}, one can get
\begin{equation}
\frac{\widetilde{\Delta}^{}_{21} \widetilde{\Delta}^{}_{31} \widetilde{\Delta}^{}_{32}}{\Delta^{}_{21} \Delta^{}_{31} \Delta^{}_{32}} \approx \widehat{C}^{}_{12} \widehat{C}^{}_{13}  \; ,
\label{eq:naumov}
\end{equation}
which is not obtainable directly from the expressions of $\Delta^{}_{ij}$ in Eqs.~(\ref{eq:Del21rev})-(\ref{eq:Del32rev}). From Eq.~(\ref{eq:jratio}), it is then evident that the Jarlskog invariant in matter in $\widetilde{\cal J}$ is determined by $\widehat{C}^{}_{12} \widehat{C}^{}_{13}$, which will be dramatically modified by the resonances at $A^{}_{\rm c} = \alpha^{}_{\rm c} \cos 2\theta^{}_{12}$ and $A^{}_{\rm c} = \cos 2\theta^{}_{13}$ in addition to the overall suppression for increasing $A^{}_{\rm c}$. In Fig.~\ref{fig:J}, the analytical result in Eq.~(\ref{eq:jratio}) has been plotted along with the exact numerical result, showing an excellent agreement.

Finally, let us make some comments on the analytical formulas of the absolute square of the effective mixing matrix element $|V^{}_{\alpha i}|^2_{}$ for $\alpha = e, \mu, \tau$ and $i = 1, 2, 3$. Given three mixing angles and the CP-violating phase in matter, it is in principle straightforward to reconstruct $|V^{}_{\alpha i}|^2$. For example, we have $|V^{}_{e1}|^2 = \cos^2 \widetilde{\theta}^{}_{13} \cos^2 \widetilde{\theta}^{}_{12}$, where the approximate formulas of $\cos^2 \widetilde{\theta}^{}_{13}$ and $\cos^2 \widetilde{\theta}^{}_{12}$ have been given in Eq.~(\ref{eq:solth13no}) and Eq.~(\ref{eq:solth12}), respectively. With the help of these two equations, one then arrives at
\begin{eqnarray}
|V^{}_{e 1}|^{2}_{} &=& \frac{\cos^{2}_{}\theta^{}_{13}}{2}\left( 1-\frac{A^{}_{\ast}-\cos 2 \theta^{}_{12}}{\widehat{C}^{}_{12}} \right) \; . \label{eq:Ve1}
	\end{eqnarray}
In the limit that the matter effects are negligible (i.e., $A^{}_{\rm c} \ll \alpha^{}_c$ or $A^{}_* = a/\Delta^{}_{21} \ll 1$), one can observe that $\widetilde{\Delta}^{}_{21} \approx \Delta^{}_{\rm c}\widehat{C}^{}_{12} \alpha^{}_{\rm c}$ is a good approximation. Using the measured values of neutrino mixing angles $\cos^{2}_{}\theta^{}_{13} \approx 1$ and $\cos^{2}_{}\theta^{}_{12} \approx 2/3$, we can verify that Eq.~(\ref{eq:Ve1}) can be reduced to
\begin{eqnarray}
|V^{}_{e1}|^{2}_{} \approx \frac{1}{2}\left( 1- \frac{a - \Delta^{}_{21}/3}{\widetilde{\Delta}^{}_{21} }\right) \; , 
\label{eq:Ve1d}
\end{eqnarray}
which is just Eq.~(39) in Ref.~\cite{Chiu:2017ckv}. The results of other matrix elements $|V^{}_{e2}|^2$, $|V^{}_{e3}|^2$, $|V^{}_{\mu 3}|^2$ and $|V^{}_{\tau 3}|^2$ can be computed in a similar way.

However, as is well known, the explicit expressions of $|V^{}_{\mu 1}|^2_{}$, $|V^{}_{\mu 2}|^2_{}$, $|V^{}_{\tau 1}|^2_{}$ and $|V^{}_{\tau 2}|^2_{}$ should be very complicated in the standard parametrization. With the help of Eq.~(\ref{eq:jratio}), we find that the final results can be simplified to a large extent. For illustration, we obtain
\begin{eqnarray}
|V^{}_{\mu 1}|^{2}_{} \approx \cos^{2}_{}\theta^{}_{23}-\left(\sin^{2}_{}\theta^{}_{23}+\frac{\cos 2\theta^{}_{23}}{\cos^2_{}\widetilde{\theta}^{}_{13}}\right)|V^{}_{e1}|^{2}_{}-\frac{2 \widehat{C}^{}_{13}{\cal J}\cot \delta}{\widehat{C}^{}_{12}\cos^2_{}\widetilde{\theta}^{}_{13}} \; ,
\label{eq:Vmu1}
\end{eqnarray}
where the zeroth-order relations $\widetilde{\theta}^{}_{23} \approx \theta^{}_{23}$ and $\widetilde{\delta} \approx \delta$ have been used. The other three elements can be examined similarly. It is worthwhile to notice that the formulas in Eqs.~(\ref{eq:Ve1}) and (\ref{eq:Vmu1}) are more complicated but valid in a broader range of $A^{}_{\rm c}$, compared to their counterparts in Ref.~\cite{Chiu:2017ckv}.
	
\subsection{Further Discussions}

In this subsection, we discuss the analytical solutions to the effective neutrino masses and mixing parameters in the IO case. As in the NO case, we define $A^{}_{\rm c} \equiv a/\Delta^{}_{\rm c}$ and $\alpha^{}_{\rm c} \equiv \Delta^{}_{21}/\Delta^{}_{\rm c} < 0$. 

\begin{figure}[t!]
	\centering
	\includegraphics[width=0.49\textwidth]{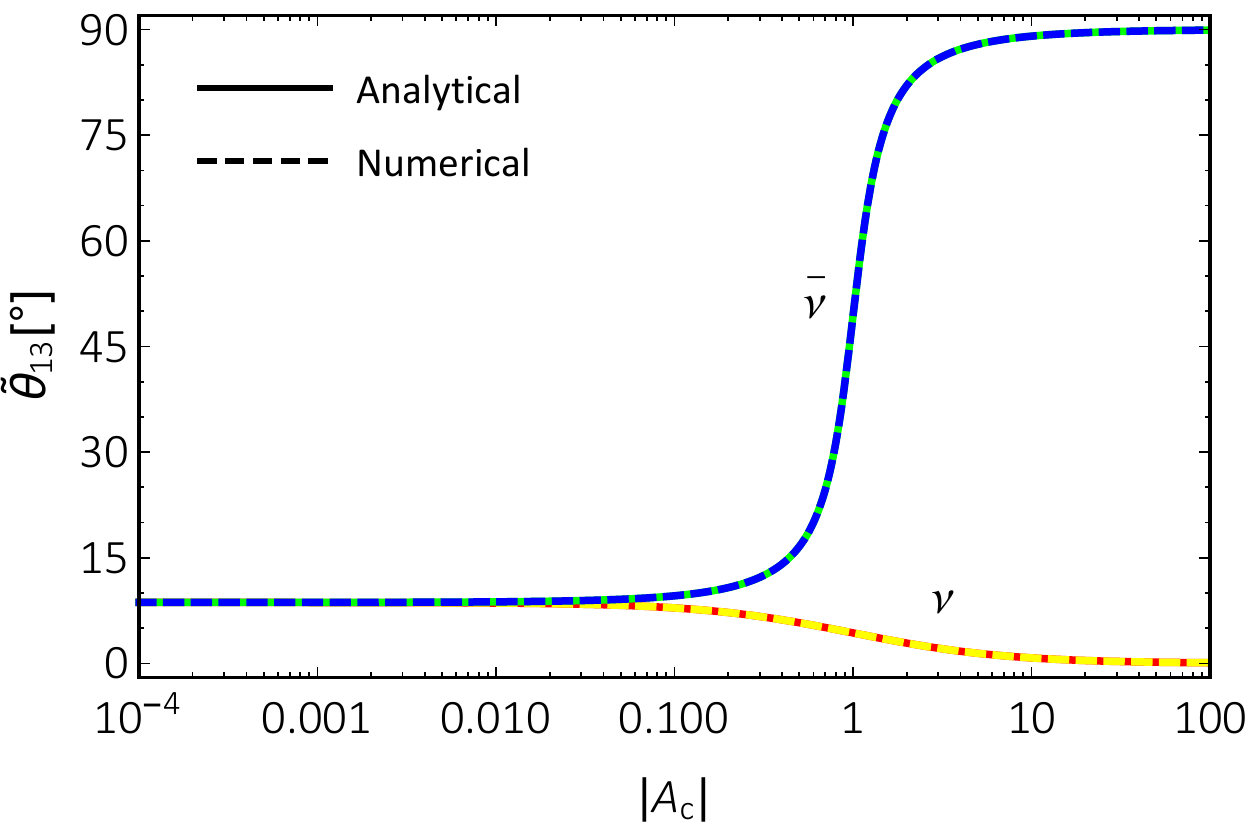}
	\includegraphics[width=0.49\textwidth]{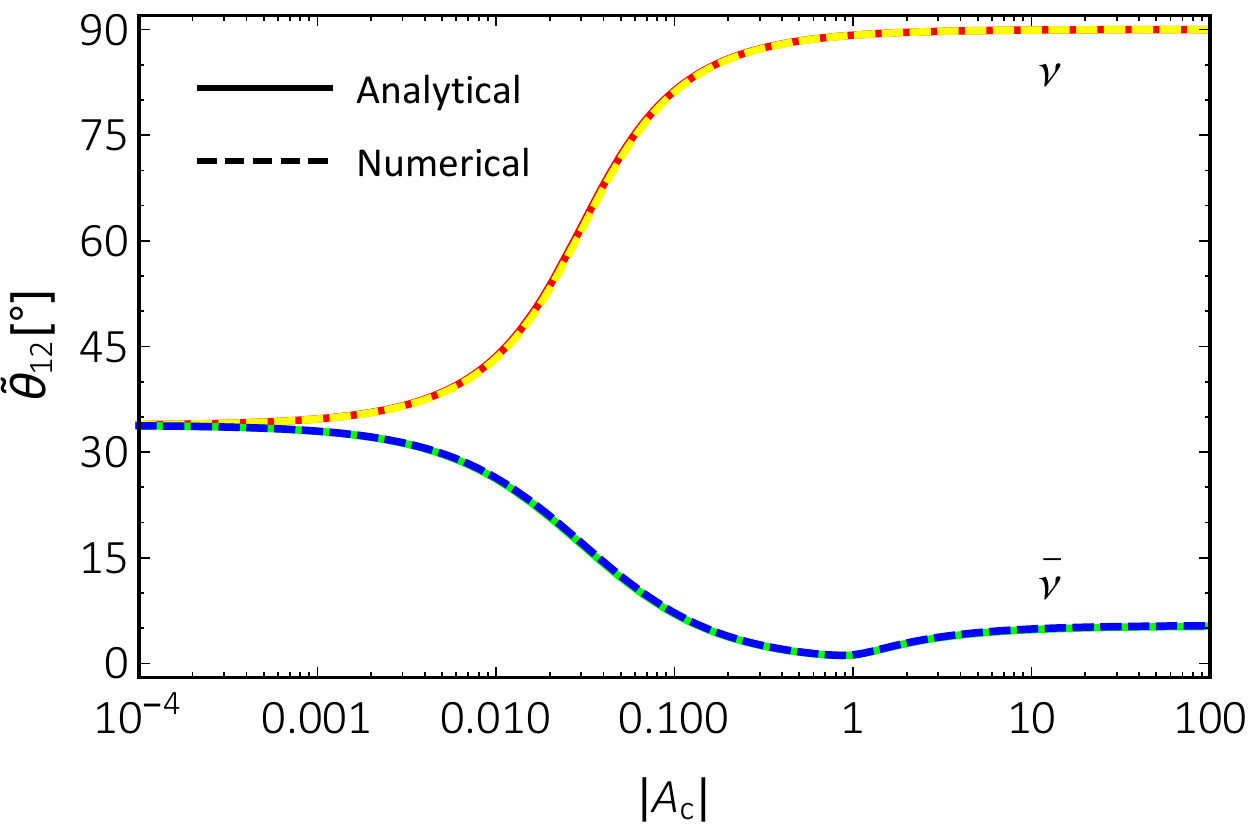}
	\vspace{-0.2cm}
	\caption{Analytical and numerical solutions to $\widetilde{\theta}^{}_{13}$ ({\it left panel}) and $\widetilde{\theta}_{12}$ ({\it right panel}) in the case of inverted neutrino mass ordering, where the best-fit values of neutrino mixing parameters from Ref.~\cite{Esteban:2018azc} have been input and the conventions for the curves are the same as in the previous figures.}
	\label{fig:Ith1312} 
\end{figure}

It is convenient to start with the effective neutrino mass-squared differences for neutrinos to the first order of $\alpha^{}_{\rm c}$, namely,
\begin{eqnarray}
\widetilde{\Delta}^{}_{21} &\approx& -\Delta^{}_{\rm c} \left[\frac{1}{2}\left(1 + A^{}_{\rm c} - \widehat{C}^{}_{13}\right) - \alpha^{}_{\rm c} \cos 2\theta^{}_{12} \right] \; , \label{eq:IDel21c} \\
\widetilde{\Delta}^{}_{31} &\approx& +\Delta^{}_{\rm c} \widehat{C}^{}_{13} \; , \label{eq:IDel31c} \\
\widetilde{\Delta}^{}_{32} &\approx& +\Delta^{}_{\rm c} \left[\frac{1}{2}\left(1 + A^{}_{\rm c} + \widehat{C}^{}_{13}\right) - \alpha^{}_{\rm c} \cos 2\theta^{}_{12} \right] \; , \label{eq:IDel32c}
\end{eqnarray}
which in comparison with Eqs. (\ref{eq:Del21c})-(\ref{eq:Del32c}) reveal that the expressions of $\widetilde{m}_1^2$ and $\widetilde{m}_2^2$ have been exchanged. Such an arrangement of three effective neutrino mass eigenvalues in matter guarantees $\widetilde{m}^2_3 < \widetilde{m}^2_1 < \widetilde{m}^2_2$ for $|A_{\rm c}| \to \infty$. This is also in accordance with the convention for Eqs. (\ref{eq:Del21c})-(\ref{eq:Del32c}), where $\widetilde{m}^2_1 < \widetilde{m}^2_2 < \widetilde{m}^2_3$ for $A^{}_{\rm c} \to \infty$ is satisfied in the NO case. With the help of Eqs. (\ref{eq:IDel21c})-(\ref{eq:IDel32c}), we can follow the same procedure leading to Eqs. (\ref{eq:solth13a}) and (\ref{eq:solth13b}) and then obtain that the solution to $\cos^2 \widetilde{\theta}^{}_{13}$ is still given by Eq.~(\ref{eq:solth13no}).
For antineutrinos, the result can be derived by replacing $A^{}_{\rm c}$ in Eq.~(\ref{eq:solth13no}) with $-A^{}_{\rm c}$. It should be noticed that $A^{}_{\rm c} \equiv a/\Delta^{}_{\rm c}$ is negative for neutrinos in the IO case, and the analytical results for antineutrinos in the same case have been derived by using the same definition of $A^{}_{\rm c}$. Therefore, it is convenient to plot all the effective mixing parameters and neutrino mass-squared differences with respect to the absolute value $|A^{}_{\rm c}|$. In the left panel of Fig.~\ref{fig:Ith1312}, the analytical and numerical solutions to $\widetilde{\theta}^{}_{13}$ have been presented for both neutrinos and antineutrinos. In our numerical calculations, the best-fit values of neutrino mixing parameters $\{\theta^{}_{12} = 33.82^{\circ},  \theta^{}_{13} = 8.65^{\circ}, \theta^{}_{23} = 49.8^{\circ}, \delta = 284^{\circ}\}$ and neutrino mass-squared differences $\{\Delta m^{2}_{21} = 7.39 \times 10^{-5}~{\rm eV^2}, \Delta m^{2}_{32} = - 2.512\times 10^{-3}~{\rm eV^2}\}$ from Ref.~\cite{Esteban:2018azc} have been used. Comparing the left panels of Fig.~\ref{fig:Ith1312} and Fig.~\ref{fig:th13}, one can realize that the evolution of $\widetilde{\theta}^{}_{13}$ for antineutrinos in the IO case is exactly the same as that for neutrinos in the NO case. This can be easily understood in the two-flavor neutrino mixing limit, for which the relevant mixing angle in vacuum is $\theta^{}_{13}$ and the mass-squared difference in vacuum is $\Delta^{}_{\rm c}$. When $A^{}_{\rm c}$ is positive, which is true for neutrinos in the NO case and for antineutrinos in the IO case, the resonance condition $A^{}_{\rm c} = \cos 2\theta^{}_{13}$ can be fulfilled. In other cases, where $A^{}_{\rm c}$ turns out to be negative, there will be no resonance and the matter effects tend to suppress the effective mixing angle $\widetilde{\theta}^{}_{13}$. For comparison, we have listed the analytical results of the effective parameters for neutrinos and antineutrinos in both NO and IO cases in Table~\ref{tab:expr}.

Next, we will continue with the solution to $\widetilde{\theta}^{}_{12}$. As in the NO case, two auxiliary functions ${\cal F}(A^{}_{\rm c})$ and ${\cal G}(A^{}_{\rm c})$ are also introduced to modify the effective neutrino mass-squared differences, namely,
\begin{eqnarray}
\widetilde{\Delta}^{}_{21} &\approx& - \Delta^{}_{\rm c} \left[\frac{1}{2}\left(1 + A^{}_{\rm c} - \widehat{C}^{}_{13}\right) + \alpha^{}_{\rm c} ({\cal F} - {\cal G}) \right] \; , \label{eq:IDel21n} \\
\widetilde{\Delta}^{}_{31} &\approx& + \Delta^{}_{\rm c} \left( \widehat{C}^{}_{13} + \alpha^{}_{\rm c} {\cal G}\right) \; , \label{eq:IDel31n} \\
\widetilde{\Delta}^{}_{32} &\approx& + \Delta^{}_{\rm c} \left[\frac{1}{2}\left(1 + A^{}_{\rm c} + \widehat{C}^{}_{13}\right) + \alpha^{}_{\rm c} {\cal F} \right] \; . \label{eq:IDel32n}
\end{eqnarray}
To correctly reproduce the neutrino mass-squared differences at $A^{}_{\rm c} = 0$ and maintain the result of $\widetilde{\theta}^{}_{13}$, we require that the conditions ${\cal F} + {\cal G} = -\cos 2\theta^{}_{12}$, ${\cal F}(0) = \sin^2 \theta^{}_{12}$ and ${\rm d}{\cal F}/{\rm d}A^{}_{\rm c} + {\rm d}{\cal G}/{\rm d}A_{\rm c}=0$ should be satisfied. In the same way as in the NO case, it is straightforward to get
\begin{equation}
\sin^2\widetilde{\theta}_{12}=\frac{1}{2}\left(1+\frac{A_*-\cos2\theta_{12}}{\widehat{C}_{12}^{}}\right) \; ,
\label{eq:Itheta12sol}
\end{equation}
where  $A^{}_* \equiv A^{}_{\rm c}/\alpha^{}_{\rm c} = a/\Delta^{}_{21}$ and $\widehat{C}^{}_{12} \equiv \sqrt{1 - 2A^{}_* \cos 2\theta^{}_{12} + A^2_*}$ have been defined as before. For antineutrinos, since the resonance at $A^{}_{\rm c} = \cos 2\theta^{}_{13}$ will greatly change $\widetilde{\theta}^{}_{13}$, a correction factor should be included to take account of the resonant enhancement. Consequently, we have
\begin{equation}
\sin^2 \widetilde{\theta}^{}_{12} =  \frac{1}{2} \left(1 - \frac{A^{}_* + \cos 2\theta^{}_{12}}{\widehat{C}^{\prime}_{12}}\right) \frac{\cos^2\theta_{13}}{\cos^2\widetilde{\theta}_{13}} \; ,
\label{eq:Isolth12}
\end{equation}
where $\widehat{C}_{12}^{\prime} \equiv \sqrt{1 + 2A^{}_* \cos 2\theta^{}_{12} + A_*^2}$ is just $\widehat{C}^{}_{12}$ with $A^{}_*$ replaced by $-A^{}_*$. The analytical result in Eq.~(\ref{eq:Itheta12sol}) for neutrinos and that in Eq.~(\ref{eq:Isolth12}) for antineutrinos have been depicted in the right panel of Fig.~\ref{fig:Ith1312}, together with the exact numerical results. The analytical results agree well with the numerical ones, and the differences between these two results should be on the same order as those in the NO case. For $\widetilde{\theta}^{}_{12}$, the main difference between the results in the NO and IO cases is that the resonance at $A^{}_{\rm c} = \cos 2\theta^{}_{13}$ occurs in the neutrino and antineutrino sector, respectively.

\begin{figure}[t!]
	\centering
	\includegraphics[width=0.49\textwidth]{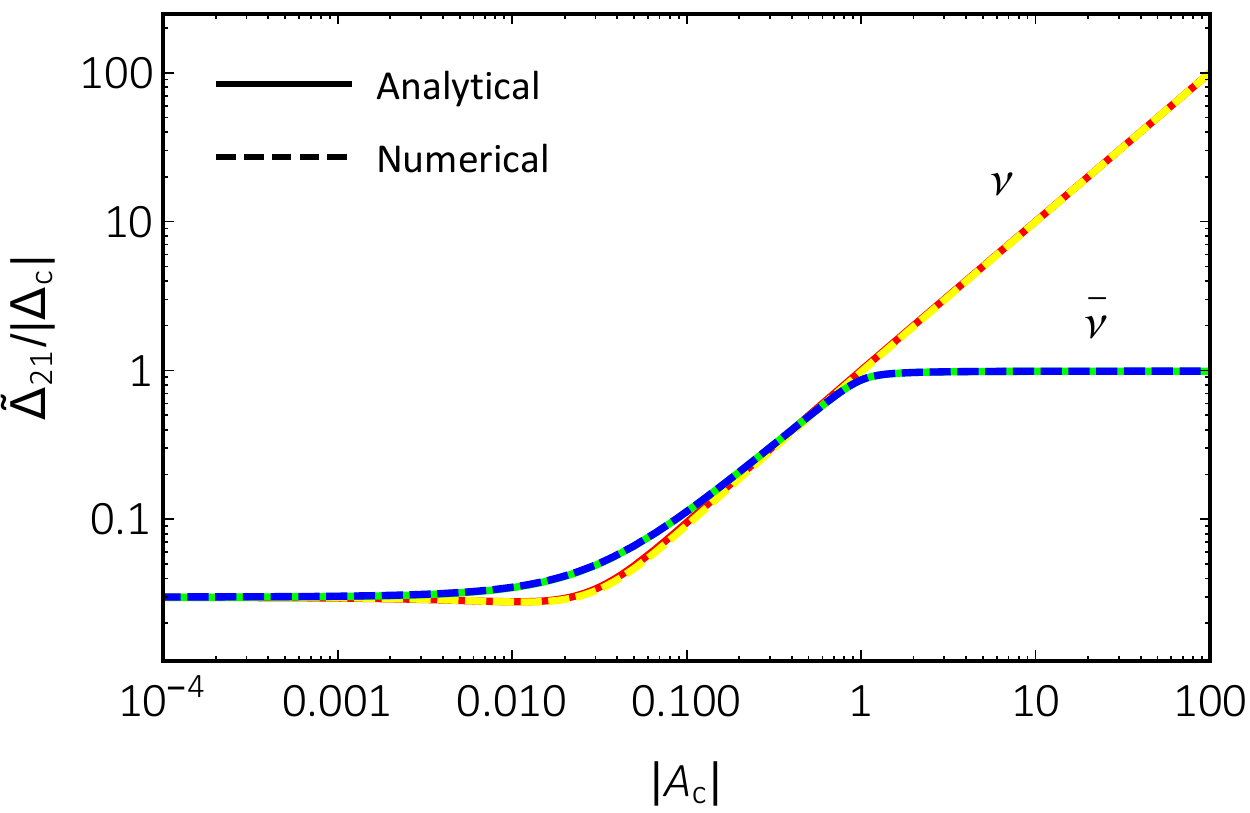}
	\includegraphics[width=0.49\textwidth]{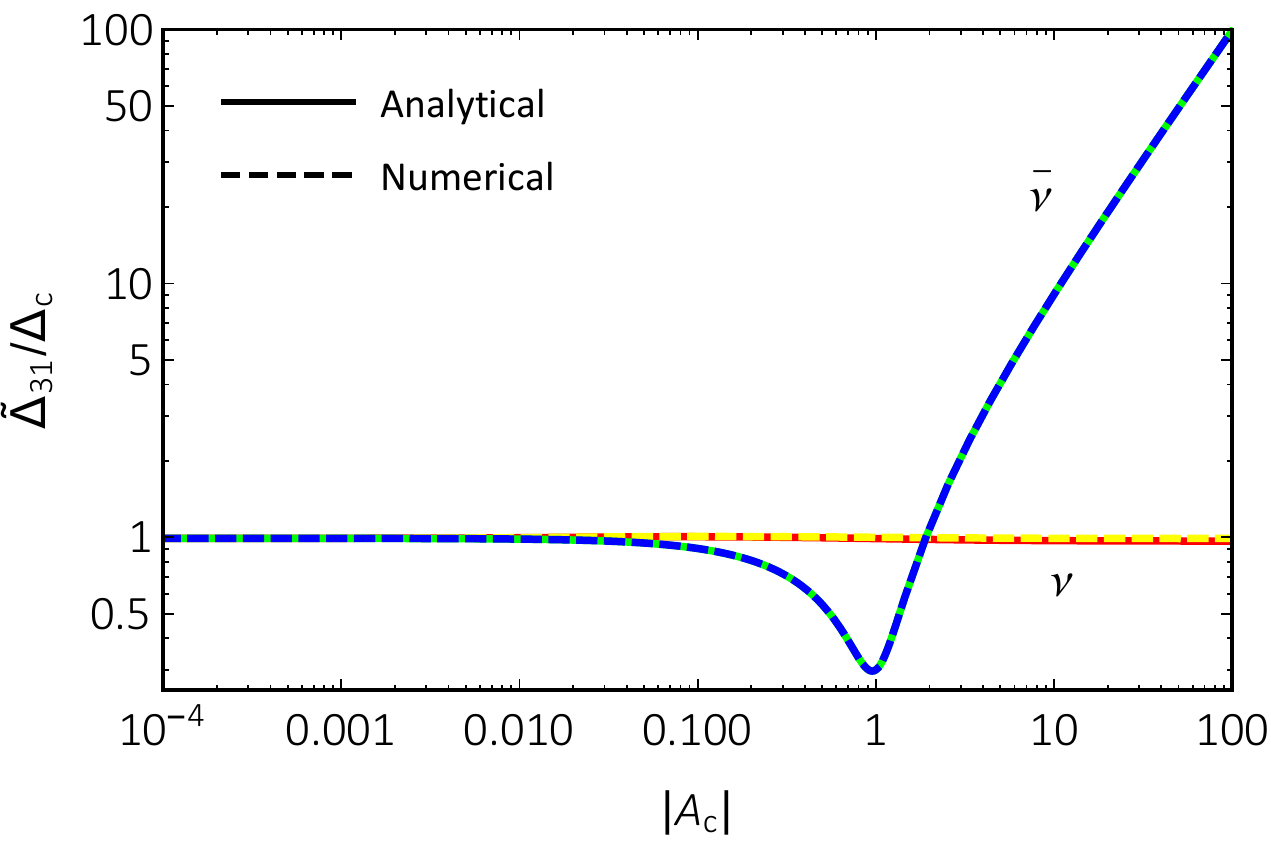}
	\includegraphics[width=0.49\textwidth]{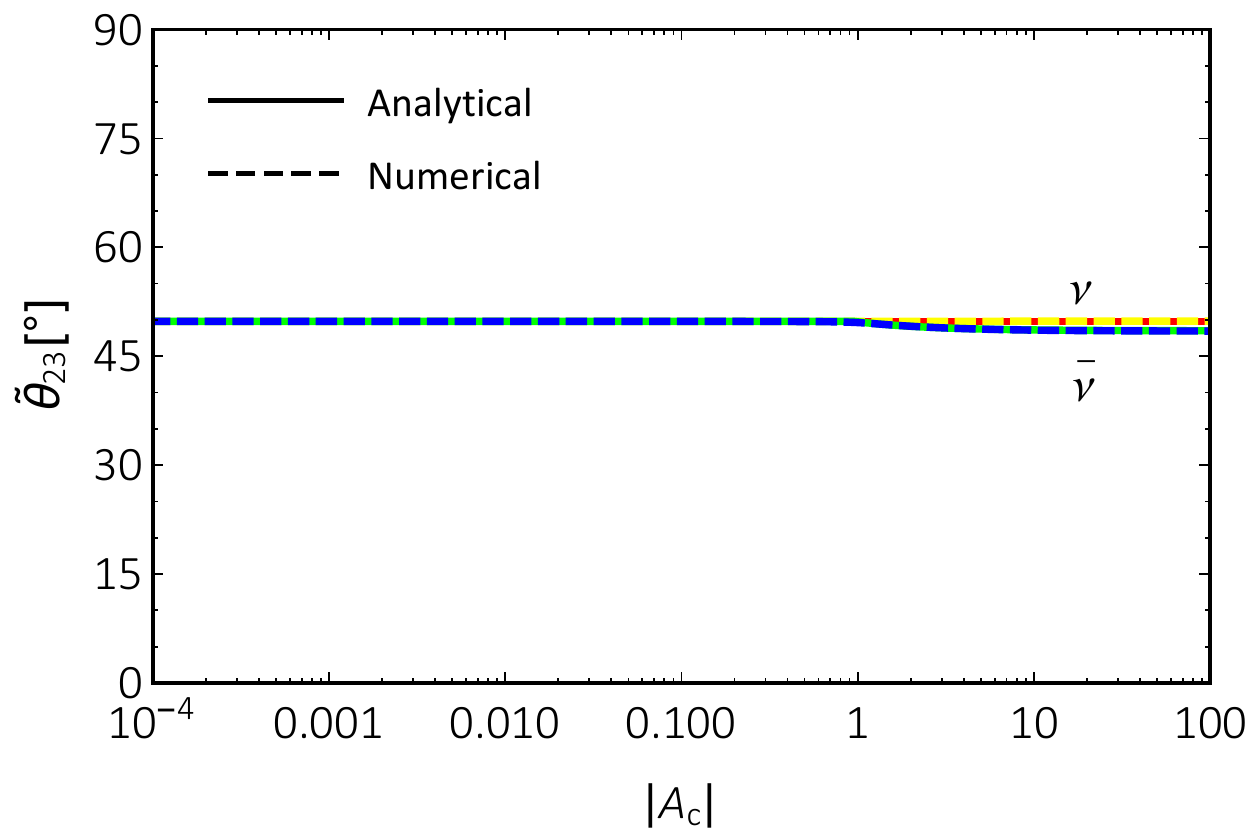}
	\includegraphics[width=0.495\textwidth]{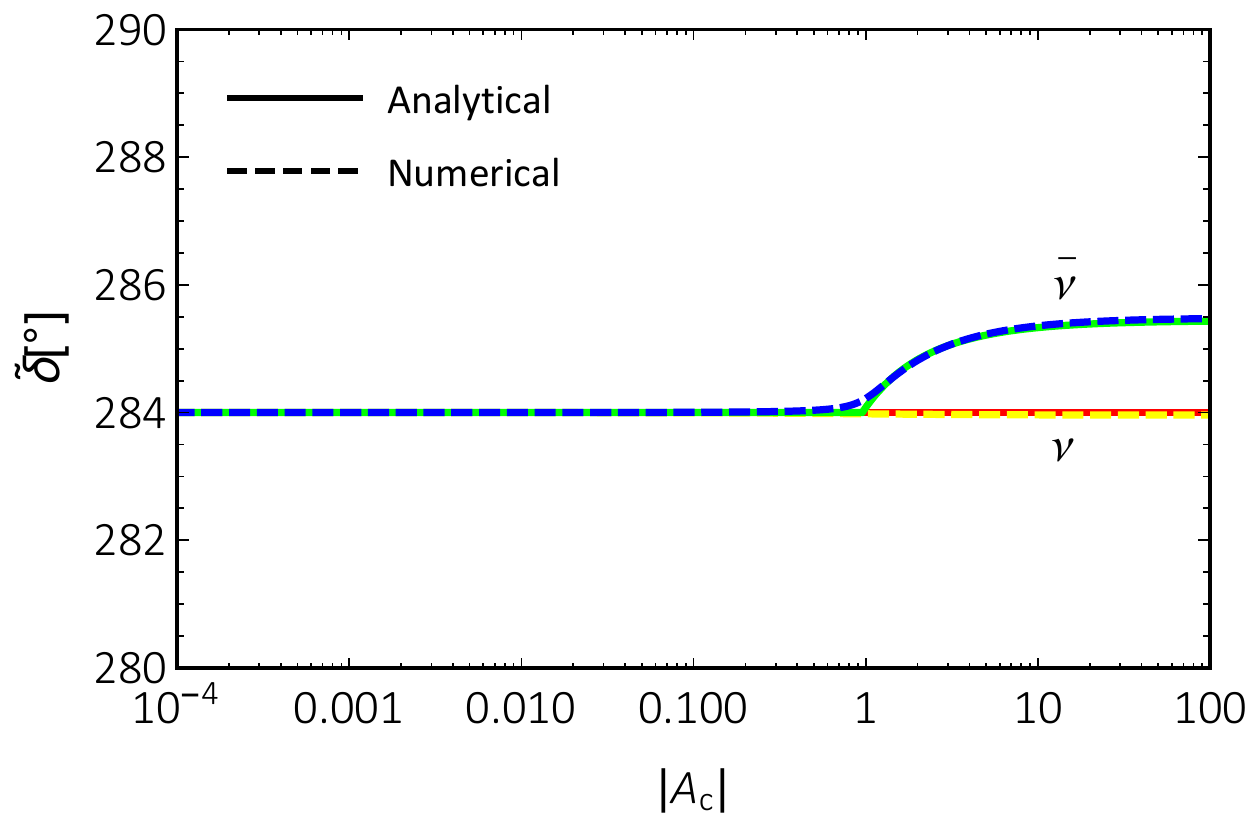}
	\vspace{-0.2cm}
	\caption{Analytical and numerical solutions to $\{\widetilde{\Delta}^{}_{21}/|\Delta^{}_{\rm c}|, \widetilde{\Delta}^{}_{31}/\Delta^{}_{\rm c}\}$ are shown in the first row while those to $\{\widetilde{\theta}^{}_{23}, \widetilde{\delta}\}$ in the second row, where the inverted neutrino mass ordering is assumed and the input parameters and conventions for the curves are the same as in the previous figures.}
	\label{fig:Alltheother} 
\end{figure}

\begin{figure}[t!]
	\centering
	\includegraphics[width=0.49\textwidth]{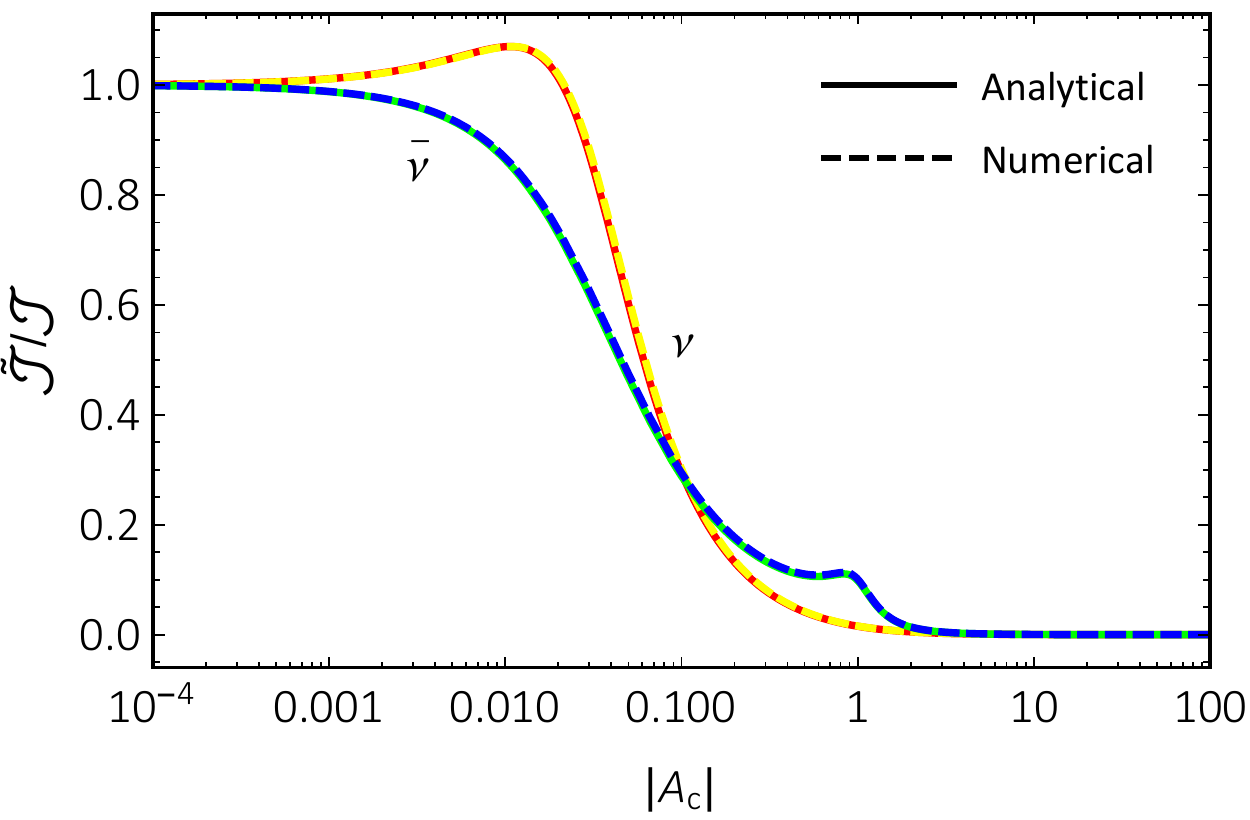}
	\vspace{-0.2cm}
	\caption{Analytical and numerical solutions to $\widetilde{\cal J}/{\cal J}$ in the inverted mass ordering case, where the input parameters and conventions for the curves are the same as before.}
	\label{fig:IJ} 
\end{figure}

As the auxiliary functions ${\cal F}(A^{}_{\rm c})$ and ${\cal G}(A^{}_{\rm c})$ have been fixed, three effective neutrino mass-squared differences are then found to be
\begin{eqnarray}
\widetilde{\Delta}_{21} &=& -\Delta^{}_{\rm c} \left[\frac{1}{2}(1 + A^{}_{\rm c} - \widehat{C}_{13}^{}) - (\widehat{C}^{}_{12} + A^{}_*) \alpha^{}_{\rm c}\right] \; , \label{eq:IDel21rev}\\
\widetilde{\Delta}^{}_{31} &=& + \Delta^{}_{\rm c} \left[\widehat{C}_{13}^{} + \frac{1}{2}\left(\widehat{C}^{}_{12} + A^{}_* - \cos2\theta^{}_{12}\right) \alpha^{}_{\rm c}\right] \; , \label{eq:IDel31rev} \\
\widetilde{\Delta}^{}_{32} &=& + \Delta^{}_{\rm c} \left[\frac{1}{2}(1 + A^{}_{\rm c} + \widehat{C}_{13}^{}) - \frac{1}{2}\left(\widehat{C}^{}_{12} + A^{}_* + \cos2\theta^{}_{12}\right) \alpha^{}_{\rm c}\right] \; , \label{eq:IDel32rev}
\end{eqnarray}
for neutrinos. The results for antineutrinos can be derived by replacing $A^{}_{\rm c}$ with $-A^{}_{\rm c}$ and $A^{}_{*}$ with $-A^{}_{*}$ in the above equations. In addition, $\widetilde{\theta}^{}_{23} \approx \theta^{}_{23}$ and $\widetilde{\delta} \approx \delta$ hold excellently for neutrinos. However, for antineutrinos, we have $\widetilde{\theta}^{}_{23} = \theta^{}_{23}$ if $A^{}_{\rm c} \leq \cos 2\theta^{}_{13}$, otherwise $\widetilde{\theta}^{}_{23}$ is given by the same formula in Eq.~(\ref{eq:solth23}). The effective CP-violating phase $\widetilde{\delta}$ can be calculated by using the Toshev relation $\sin \widetilde{\delta} = \sin 2\theta^{}_{23} \sin\delta/\sin 2\widetilde{\theta}^{}_{23}$. Finally, the ratio of the effective Jarlskog invariant to its counterpart in vacuum is $\widetilde{\cal J}/{\cal J} = 1/(\widehat{C}^{}_{12} \widehat{C}^{}_{13})$ in the neutrino case. Although the formulas in the IO case take the same form as in the NO case, it is worthwhile to mention that the structure of resonances
can be very different. The analytical and numerical solutions to $\widetilde{\Delta}^{}_{21}/|\Delta^{}_{\rm c}|$ and $\widetilde{\Delta}^{}_{31}/\Delta^{}_{\rm c}$, as well as those to $\widetilde{\delta}^{}_{23}$ and $\widetilde{\delta}$, are plotted in Fig.~\ref{fig:Alltheother}. Meanwhile, the result for $\widetilde{\cal J}/{\cal J}$ in the IO case is shown in Fig.~\ref{fig:IJ}, where one can easily recognize the single resonance at $A^{}_{\rm c} = \alpha^{}_{\rm c} \cos 2\theta^{}_{12}$ for neutrinos, and that at $A^{}_{\rm c} = \cos 2\theta^{}_{13}$ for antineutrinos. This observation should be compared with the resonance structure in the NO case, where two resonances appear in the neutrino sector while no resonance in the antineutrino sector.

\section{Concluding Remarks}

In light of the ongoing and forthcoming long-baseline accelerator neutrino experiments and huge atmospheric neutrino observatories, it will be very helpful to have a better understanding of matter effects on neutrino oscillations. Adopting the language of the renormalization-group equations (RGEs) for the effective neutrino masses and mixing parameters~\cite{Xing:2018}, we show in this paper that analytical solutions to the RGEs can be derived with the help of series expansion of neutrino mass eigenvalues in matter~\cite{Freund:2001, Akhmedov2004}. The essential idea is to regularize the series expansion in the region where it is invalid by the exact RGEs. Some interesting observations have been made and summarized as follows.

We take neutrino oscillations in matter in the NO case for example. First, the effective mixing angle $\widetilde{\theta}^{}_{13}$ can be excellently described by the formula in Eq.~(\ref{eq:solth13no}) in the limit of two-flavor neutrino mixing~\cite{Freund:2001}. This is also true for the effective mixing angle $\widetilde{\theta}^{}_{12}$ except in the region of large matter effects (e.g., $A^{}_{\rm c} \gtrsim \cos 2\theta^{}_{13}$), where a correction factor $\cos^2 \theta^{}_{13}/\cos^2 \widetilde{\theta}^{}_{13}$ should be included. Second, it is widely known that the matter effects can hardly modify $\widetilde{\theta}^{}_{23}$ and $\widetilde{\delta}$, so $\widetilde{\theta}^{}_{23} \approx \theta^{}_{23}$ and $\widetilde{\delta} \approx \delta$ are very good approximations until the resonance at $A^{}_{\rm c} = \cos 2\theta^{}_{13}$ is met. The deviations of $\widetilde{\theta}^{}_{23}$ and $\widetilde{\delta}$ from $\theta^{}_{23}$ and $\delta$ become significant for $A^{}_{\rm c} \gtrsim \cos 2\theta^{}_{13}$ and can be accounted for by the analytical formula in Eq.~(\ref{eq:solth23}). Finally, the Jarlskog invariant in matter is found to be $\widetilde{\cal J} \approx {\cal J}/(\widehat{C}^{}_{12} \widehat{C}^{}_{13})$, where $\widehat{C}^{}_{12} \equiv \sqrt{(A^{}_* - \cos 2\theta^{}_{12})^2 + \sin^2 2\theta^{}_{12}}$ with $A^{}_* = a/\Delta^{}_{21}$ and $\widehat{C}^{}_{13} \equiv \sqrt{(A^{}_{\rm c} - \cos 2\theta^{}_{13})^2 + \sin^2 2\theta^{}_{13}}$ well capture the main features of two resonances relevant for neutrino oscillations in matter. The analytical formulas for neutrinos and antineutrinos in both NO and IO cases are compared and listed in Table~\ref{tab:expr}.

The analytical formulas for the effective neutrino mass-squared differences $\widetilde{\Delta}^{}_{ij}$ for $ij = 21, 31, 32$ and mixing parameters $\{\widetilde{\theta}^{}_{12}, \widetilde{\theta}^{}_{13}, \widetilde{\theta}^{}_{23}, \widetilde{\delta}\}$ can be further used to calculate neutrino oscillation probabilities in matter. Since the agreement between these formulas and the exact numerical results has been found to be excellent, the implementation of analytical formulas in the phenomenological studies of neutrino oscillations will greatly increase the efficiency of numerical simulations. The investigation of neutrino oscillation probabilities in matter along this line deserves more attention and will be left for future works.

\section*{Acknowledgements}

The authors are indebted to Prof. Zhi-zhong Xing and Dr. Shao-feng Ge for helpful discussions. This work was supported in part by the National Natural Science Foundation of China under grant No.~11775232 and No.~11835013, and by the CAS Center for Excellence in Particle Physics.

\begin{landscape}
\begin{table}[]
\renewcommand\arraystretch{2.3}
\centering
\footnotesize
\begin{tabular}{c||cc|cc}
\toprule[1.2pt] \hline
& \multicolumn{2}{c|}{Normal Mass Ordering} & \multicolumn{2}{c}{Inverted Mass Ordering} \\ \cline{2-5}
			&    Neutrino       &   Antineutrino       &     Neutrino      &    Antineutrino       \\ \hline
			$\cos^2 \widetilde{\theta}^{}_{13}$		&   $\displaystyle \frac{1}{2} \left(1 - \frac{A^{}_{\rm c} - \cos 2\theta^{}_{13}}{\widehat{C}^{}_{13}} \right)$      &    $\displaystyle \frac{1}{2} \left( 1 + \frac{A^{}_{\rm c} + \cos 2\theta^{}_{13}}{\widehat{C}^{\prime}_{13}} \right)$       &    $\displaystyle \frac{1}{2} \left( 1 - \frac{A^{}_{\rm c} - \cos 2\theta^{}_{13}}{\widehat{C}^{}_{13}} \right)$       &    $\displaystyle \frac{1}{2} \left( 1 + \frac{A^{}_{\rm c} + \cos 2\theta^{}_{13}}{\widehat{C}^{\prime}_{13}} \right)$       \\ 
			$\cos^2 \widetilde{\theta}^{}_{12}$	&    $\displaystyle \frac{1}{2} \left(1 - \frac{A^{}_* - \cos 2\theta^{}_{12}}{\widehat{C}^{}_{12}}\right) \frac{\cos^2 \theta^{}_{13}}{\cos^2 \widetilde{\theta}^{}_{13}}$      &    $\displaystyle \frac{1}{2} \left(1 + \frac{A^{}_* + \cos 2\theta^{}_{12}}{\widehat{C}^{\prime}_{12}}\right)$       &    $\displaystyle \frac{1}{2} \left(1 - \frac{A^{}_* - \cos 2\theta^{}_{12}}{\widehat{C}^{}_{12}}\right)$       &    $\displaystyle 1 - \frac{1}{2} \left(1 - \frac{A^{}_* + \cos 2\theta^{}_{12}}{\widehat{C}^{\prime}_{12}}\right) \frac{\cos^2\theta_{13}}{\cos^2\widetilde{\theta}_{13}}$            \\
			$\widetilde{\theta}^{}_{23}$		&   $\begin{cases} \theta^{}_{23} \; , ~&  A^{}_{\rm c} \leq \cos 2\theta^{}_{13} \\ \theta^{}_{23} + \Delta \theta^{}_{23} \; , ~& A^{}_{\rm c} > \cos 2\theta^{}_{13} \end{cases} $        &     $\theta^{}_{23}$      &     $\theta^{}_{23}$      &     $\begin{cases} \theta^{}_{23} \; , ~&  A^{}_{\rm c} \leq \cos 2\theta^{}_{13} \\ \theta^{}_{23} + \Delta \theta^{}_{23} \; , ~& A^{}_{\rm c} > \cos 2\theta^{}_{13} \end{cases}$   \\
			$\displaystyle \frac{\widetilde{\Delta}^{}_{21}}{\Delta^{}_{\rm c}}$	&  $\displaystyle \frac{1 + A^{}_{\rm c}- \widehat{C}^{}_{13}}{2}+ (\widehat{C}^{}_{12} - A^{}_*) \alpha^{}_{\rm c}$           &  $\displaystyle \frac{1 - A^{}_{\rm c}- \widehat{C}^{\prime}_{13}}{2} + (\widehat{C}^{\prime}_{12} + A^{}_*) \alpha^{}_{\rm c}$         &   $\displaystyle (\widehat{C}^{}_{12} + A^{}_* ) \alpha^{}_{\rm c} - \frac{1 + A^{}_{\rm c}-\widehat{C}_{13}^{}}{2}$       &   $\displaystyle (\widehat{C}^{\prime}_{12} - A^{}_*) \alpha^{}_{\rm c} -\frac{1 - A^{}_{\rm c} - \widehat{C}_{13}^{\prime}}{2}$        \\
			$\displaystyle \frac{\widetilde{\Delta}^{}_{31}}{\Delta^{}_{\rm c}}$		&    $\displaystyle \frac{1 + A^{}_{\rm c} + \widehat{C}^{}_{13}}{2}+ \frac{\widehat{C}^{}_{12} - A^{}_* - \cos 2\theta_{12}}{2}\alpha^{}_{\rm c}$      &    $\displaystyle \frac{1 - A^{}_{\rm c} + \widehat{C}^{\prime}_{13}}{2}+ \frac{\widehat{C}^{\prime}_{12} + A^{}_* - \cos 2\theta_{12}}{2}\alpha^{}_{\rm c}$       &  $\displaystyle \widehat{C}_{13}^{} + \frac{\widehat{C}^{}_{12} + A^{}_* - \cos 2\theta_{12}}{2} \alpha^{}_{\rm c}$         &    $\displaystyle \widehat{C}_{13}^{\prime} + \frac{\widehat{C}^{\prime}_{12} - A^{}_* - \cos 2\theta^{}_{12}}{2} \alpha^{}_{\rm c}$                   \\ 
			$\displaystyle \widetilde{\cal J}/{\cal J}$ &  $\displaystyle 1/(\widehat{C}^{}_{12} \widehat{C}^{}_{13})$ &  $\displaystyle 1/(\widehat{C}^{\prime}_{12} \widehat{C}^{\prime}_{13})$ & $\displaystyle 1/(\widehat{C}^{}_{12} \widehat{C}^{}_{13})$ &  $\displaystyle 1/(\widehat{C}^{\prime}_{12} \widehat{C}^{\prime}_{13})$ \\
\hline 
\bottomrule[1.2pt]
\end{tabular}
\caption{The analytical formulas of three effective neutrino mixing angles $\{\cos^2 \widetilde{\theta}^{}_{13},\cos^2 \widetilde{\theta}^{}_{12}, \widetilde{\theta}^{}_{23}\}$, two independent neutrino mass-squared differences $\{\widetilde{\Delta}^{}_{21}/\Delta^{}_{\rm c}, \widetilde{\Delta}^{}_{31}/\Delta^{}_{\rm c}\}$ and the Jarsklog invariant $\widetilde{\cal J}/{\cal J}$ for neutrinos and antineutrinos in both the NO and IO cases, while the effective CP-violating phase $\widetilde{\delta}$ can be calculated by using the Toshev relation $\sin \widetilde{\delta} = \sin 2\theta^{}_{23} \sin \delta/\sin 2\widetilde{\theta}^{}_{23}$ or extracted from the Jarlskog invariant $\widetilde{\cal J}$. Note that $C_{13}^{\prime} \equiv \sqrt{1 + 2 A^{}_{\rm c}\cos 2\theta^{}_{13} + A^2_{\rm c}}$, $C_{12}^{\prime} \equiv \sqrt{1 + 2 A^{}_{*} \cos 2\theta^{}_{12} + A_{*}^2}$ and $\Delta \theta^{}_{23} \equiv \alpha^{}_{\rm c} \cos \delta\sin2\theta^{}_{12} (\cos^2\theta^{}_{13} - \alpha^{}_{\rm c} \cos2\theta^{}_{12})(A^{}_{\rm c} - \cos2\theta^{}_{13}) / (2A^{}_{\rm c} \cos2\theta^{}_{13} \sin\theta^{}_{13})$ have been defined. In addition, the definitions $A^{}_{\rm c} \equiv a/ \Delta^{}_{\rm c}$ and $\alpha^{}_{\rm c} \equiv \Delta^{}_{21}/\Delta^{}_{\rm c}$ are applicable in all the scenarios.}
\label{tab:expr}
\end{table}	
\end{landscape}

\end{document}